\pdfoutput=1
\documentclass[11pt,twoside,a4paper,cmspaper,final,collab]{cms-tdr}
\def\svnVersion{de9aca0}\def\svnDate{2023/06/01}\def\cmsCernNoTag{CERN-EP-2022-019}\def\cmsCernDate{\today}\def\cmsMessage{Published in the Journal of High Energy Physics as \href{http://dx.doi.org/10.1007/JHEP05(2023)233}{\doi{10.1007/JHEP05(2023)233}.}}
\begin{document}\cmsNoteHeader{HIG-19-014}

\newcommand{\LumiT}{138}
\newcommand{\Lumia}{36.3}
\newcommand{\Lumib}{41.5}
\newcommand{\Lumic}{59.8}
\newcommand{\mH}{125.38}
\newcommand{\signalstrength}{\ensuremath{2.4\pm0.9}}
\newcommand{\signalstrengthExpanded}{\ensuremath{2.4^{+0.8}_{-0.9}}\stat\,^{+0.3}_{-0.2}\syst}
\newcommand{\signalstrengthdijetthree}{\ensuremath{12.3^{+3.7}_{-3.5}}}
\newcommand{\br}{\ensuremath{0.21\pm0.08}\xspace}
\newcommand{\brExpanded}{\ensuremath{0.21^{+0.07}_{-0.08}\stat\,^{+0.03}_{-0.02}\syst}\xspace}
\newcommand{\explimit}{1.8}
\newcommand{\obslimit}{4.1}
\newcommand{\expsig}{1.2}
\newcommand{\obssig}{2.7}
\newcommand{\Pellp}{\HepParticle{\ell}{}{+}\Xspace}
\newcommand{\Pellm}{\HepParticle{\ell}{}{-}\Xspace}
\newcommand{\pp}{\ensuremath{\Pp\Pp}\xspace}
\newcommand{\Dkin}{\ensuremath{\mathcal{D_{\text{kin}}}}\xspace}
\newcommand{\DVBF}{\ensuremath{\mathcal{D}_{\text{VBF}}}\xspace}
\newcommand{\zg}{\ensuremath{\PZ\PGg}\xspace}
\newcommand{\hzg}{\ensuremath{\PH\to\zg}\xspace}
\newcommand{\brzg}{\ensuremath{\mathcal{B}(\hzg)}\xspace}
\newcommand{\hmass}{\ensuremath{m_{\PH}}\xspace}
\newcommand{\hgg}{\ensuremath{\PH\to\PGg\PGg}\xspace}
\newcommand{\brgg}{\ensuremath{\mathcal{B}(\hgg)}\xspace}
\newcommand{\lplm}{\ensuremath{\Pellp\Pellm}\xspace}
\newcommand{\epem}{\ensuremath{\Pep\Pem}\xspace}
\newcommand{\mpmm}{\ensuremath{\PGmp\PGmm}\xspace}
\newcommand{\compatibility}{1.6\xspace}
\newcommand{\brRatio}{\ensuremath{1.5^{+0.7}_{-0.6}}\xspace}
\newcommand{\brRatioCompat}{1.5\xspace}
\newcommand{\channelcompatp}{0.02\xspace}
\newcommand{\channelcompatsigma}{2.3\xspace}
\newcommand{\DRgm}{\ensuremath{\DR(\PGg,\PGm)}\xspace}
\newcommand{\DRgmRatio}{\ensuremath{\DRgm/p_{\mathrm{T}\PGg}^{2}}\xspace}
\newcommand{\Ztoll}{\ensuremath{\PZ\to\lplm}\xspace}
\newcommand{\sigppH}{\ensuremath{\sigma(\pp\to\PH)}\xspace}
\newcommand{\sigxbr}{\ensuremath{\sigppH\brzg}\xspace}
\newcommand{\zgOvergg}{\ensuremath{\brzg/\brgg}\xspace}
\newcommand{\llg}{\ensuremath{\lplm\PGg}\xspace}
\newcommand{\mll}{\ensuremath{m_{\lplm}}\xspace}
\newcommand{\ggH}{\ensuremath{\Pg\Pg\PH}\xspace}
\newcommand{\VH}{\ensuremath{\PV\PH}\xspace}
\newcommand{\ttH}{\ensuremath{\ttbar\PH}\xspace}
\newcommand{\Zgstar}{\ensuremath{\PZ/\PGg^{\ast}}\xspace}
\newcommand{\mllg}{\ensuremath{m_{\llg}}\xspace}
\newcommand{\hmm}{\ensuremath{\PH\to\mpmm}\xspace}
\newcommand{\ptneut}{\ensuremath{\sum \pt^\text{neutral}}\xspace}
\newcommand{\ptchg}{\ensuremath{\sum \pt^\text{charged}}\xspace}
\newcommand{\ptgam}{\ensuremath{\sum \pt^{\PGg}}\xspace}
\newcommand{\ptpu}{\ensuremath{\pt^{\PGm,\text{PU}}}\xspace}
\newcommand{\Ztomm}{\ensuremath{\PZ\to\mpmm}\xspace}
\newcommand{\Ztoee}{\ensuremath{\PZ\to\epem}\xspace}
\newcommand{\muR}{\ensuremath{\mu_{\mathrm{R}}}\xspace}
\newcommand{\muF}{\ensuremath{\mu_{\mathrm{F}}}\xspace}
\newcommand{\WH}{\ensuremath{\PW\PH}\xspace}
\newcommand{\ZH}{\ensuremath{\PZ\PH}\xspace}

\newlength\cmsTabSkip\setlength{\cmsTabSkip}{0.5ex}
\providecommand{\cmsTable}[1]{\resizebox{\textwidth}{!}{#1}}

\cmsNoteHeader{HIG-19-014}

\title{Search for Higgs boson decays to a \texorpdfstring{\PZ}{Z} boson and a photon in proton-proton collisions at \texorpdfstring{$\sqrt{s}=13\TeV$}{sqrt(s) = 13 TeV}}

\date{\today}

\abstract{
Results are presented from a search for the Higgs boson decay \hzg, where \Ztoll with $\Pell=\Pe$ or \PGm. The search is performed using a sample of proton-proton (\pp) collision data at a center-of-mass energy of 13\TeV, recorded by the CMS experiment at the LHC, corresponding to an integrated luminosity of \LumiT\fbinv. Events are assigned to mutually exclusive categories, which exploit differences in both event topology and kinematics of distinct Higgs production mechanisms to enhance signal sensitivity. The signal strength $\mu$, defined as the product of the cross section and the branching fraction [$\sigxbr$] relative to the standard model prediction, is extracted from a simultaneous fit to the \llg invariant mass distributions in all categories and is measured to be $\mu=2.4\pm0.9$ for a Higgs boson mass of 125.38\GeV. The statistical significance of the observed excess of events is 2.7 standard deviations. This measurement corresponds to $\sigxbr=0.21\pm0.08\unit{pb}$. The observed (expected) upper limit at 95\% confidence level on $\mu$ is 4.1\,(1.8), where the expected limit is calculated under the background-only hypothesis. The ratio of branching fractions \zgOvergg is measured to be $1.5^{+0.7}_{-0.6}$, which agrees with the standard model prediction of $0.69 \pm 0.04$ at the 1.5 standard deviation level.
}

\hypersetup{
pdfauthor={CMS Collaboration},
pdftitle={Search for Higgs boson decays to a Z boson and a photon in proton-proton collisions at sqrt(s) = 13 TeV},
pdfsubject={CMS},
pdfkeywords={CMS, Higgs}}

\maketitle

\section{Introduction}\label{sec:intro}

Since the discovery of the Higgs boson~\cite{Aad_2012,Chatrchyan_2012,CMS:2013btf} at the LHC, an extensive program of measurements~\cite{PDG2020} has been undertaken to determine its properties and couplings to different types of particles and to assess whether these properties are consistent with those predicted by the standard model (SM).
With the successful running of the LHC, large data samples of proton-proton (\pp) collisions at $\sqrt s = 13\TeV$ have been accumulated, increasing the sensitivity to Higgs boson decays with small branching fractions.
Such decays also provide probes for possible contributions arising from physics beyond the SM (BSM) and include the process \hzg~\cite{Abba96, Chen12, Htollg-FB-Sun, Passarino, Campbell_2013hz, Degrassi:2019yix, Low:2011gn,Hue:2017cph,Dedes:2019bew,Hammad:2015eca,Liu:2020nsm}.

\begin{figure*}[b]
\includegraphics[width=0.8\textwidth]{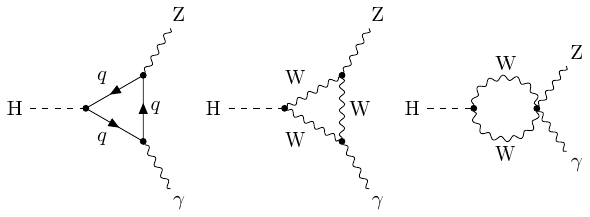}
\caption{Feynman diagrams for \hzg decay.} \label{fig:fey}
\end{figure*}

Figure~\ref{fig:fey} shows Feynman diagrams for the key SM contributions to the \hzg decay process.
Experimentally, the final state resulting from \Ztoll ($\Pell = \Pe$ or \PGm) is the most accessible, since the leptons are highly distinctive, well-measured, and provide a means to trigger the recording of the events.
In the SM, the expected branching fraction for \hzg is $\brzg = (1.57 \pm 0.09) \times 10^{-3}$, assuming a Higgs boson mass of $\hmass = \mH \pm 0.14\GeV$, taken from the most recent CMS Higgs boson mass measurement~\cite{CMS:2020xrn}.
While this branching fraction is comparable to $\brgg  = (2.27 \pm 0.04) \times 10^{-3}$~\cite{LHC-YR4,CMS:2021kom}, the \Ztoll branching fraction reduces the relative predicted signal yield.
The ratio $\zgOvergg = 0.69 \pm 0.04$ is potentially sensitive to BSM physics, such as supersymmetry and extended Higgs sectors~\cite{Djouadi:1996yq,Zg_theory_extension,Zg_theory_decaywidth,Chen:2013vi,Hung:2019jue,Archer-Smith:2020gib}.
The effects from these models can shift the \hzg and \hgg branching fractions by different amounts, making the ratio a sensitive observable.
The impact on the ratio can be $\mathcal{O}(10\%)$, depending on the model. The \hzg branching fraction is sensitive to a potential anomalous trilinear Higgs self-coupling~\cite{Degrassi:2019yix}, and a precise measurement of the branching fraction could help to test the SM prediction for this fundamental quantity.

The ATLAS and CMS Collaborations have performed searches for the decay $\hzg\to\lplm\PGg$~\cite{atl-HZG,cms-HZG,Sirunyan:2018tbk,Aad:2020plj} at $\sqrt{s}=7$, 8, and 13\TeV in the $\epem\PGg$ and $\mpmm\PGg$ final states.
The most stringent bound has been set by the ATLAS Collaboration using a data sample at $\sqrt{s} = 13\TeV$ with an integrated luminosity of 139\fbinv~\cite{Aad:2020plj}. The observed (expected) upper limit at 95\% confidence level (\CL) on \sigxbr relative to the SM is 3.6\,(2.6), assuming $\hmass=125.09\GeV$.
The region with lower dilepton invariant mass (\mll) has also been explored.
The ATLAS experiment has reported evidence for the decay $\PH\to\llg$ with $\mll < 30\GeV$ using both dilepton channels~\cite{atlas_llgrun2}.
The CMS Collaboration has also searched for the $\PH\to\llg$ process with $\mll < 50\GeV$ in the dimuon channel at $\sqrt{s}=8\TeV$~\cite{2016341} and 13\TeV~\cite{Sirunyan:2018tbk}.

This paper describes a search for the decay \hzg, where \Ztoll.
The data sample corresponds to an integrated luminosity of \LumiT\fbinv of \pp collisions at $\sqrt{s} = 13\TeV$ accumulated between 2016 and 2018.
The region at small dilepton invariant mass, $\mll < 50\GeV$, is excluded from the analysis. This region contains a contribution from an additional process, $\PH\to\PGg^{\ast}\PGg\to\llg$~\cite{Htollg-FB-Sun}.
The sensitivity of the analysis is enhanced by searching for Higgs boson production in a variety of mechanisms, including gluon-gluon fusion (\ggH); vector boson fusion (VBF); and the associated production of a Higgs boson with either a weak vector boson (\VH, where $\PV = \PZ$ or \PW) or a top quark pair (\ttH).
The dominant backgrounds arise from Drell--Yan production in association with an initial-state photon~({\Zgstar}+\PGg) and Drell--Yan production in association with jets, where a jet or additional lepton is misidentified as a photon ({\Zgstar}+jets).
After using a set of discriminating variables to suppress background in the different production mechanisms, the signal is identified as a narrow resonant peak around \hmass in the distribution of the \llg invariant mass (\mllg).

The data sample is divided into mutually exclusive categories according to (i) the presence of an additional lepton produced by $\PZ(\to\lplm)$ or $\PW(\to\Pell\PGn)$ decay, indicating the possible associated production of a Higgs boson with a \PW or \PZ boson, or \ttH production with a leptonic top quark decay; (ii) the value of a multivariate analysis (MVA) discriminant characterizing the kinematic properties of a dijet system together with the \llg candidate, indicating possible VBF production; and (iii) the value of an MVA discriminant characterizing the kinematic properties of the \llg system.
A simultaneous maximum likelihood fit is performed to the \mllg distribution in each category.
Tabulated results are provided in the HEPData record for this analysis~\cite{hepdata}.

This paper is organized as follows.
The CMS detector and event reconstruction are described in Sections~\ref{sec:cms} and~\ref{sec:reco}, respectively, and the data and simulated event samples are described in Section~\ref{sec:samples}.
Section~\ref{sec:preselection} outlines the event selection, and Section~\ref{sec:class} discusses the event categorization using the MVA discriminants described above.
The statistical procedure, including the modeling of signal and background shapes in the \mllg distributions, is presented in Section~\ref{sec:modeling}. Systematic uncertainties are discussed in Section~\ref{sec:systematics}.
The final results obtained from the fits are discussed in Section~\ref{sec:results}, followed by a summary in Section~\ref{sec:summary}.

\section{The CMS detector}\label{sec:cms}

The CMS apparatus~\cite{CMS:2008xjf} is a multipurpose, nearly hermetic detector, designed to trigger on~\cite{CMS:2020cmk,CMS:2016ngn} and identify electrons, muons, photons, and (charged and neutral) hadrons~\cite{CMS:2015xaf,CMS:2018rym,CMS:2015myp,CMS:2014pgm}.
The central feature of the CMS apparatus is a superconducting solenoid of 6\unit{m} internal diameter, providing a magnetic field of 3.8\unit{T}.
Within the solenoid volume are a silicon pixel and strip tracker, a lead tungstate crystal electromagnetic calorimeter (ECAL), and a brass and scintillator hadron calorimeter (HCAL), each composed of a barrel and two endcap sections.
The ECAL consists of 75\,848 lead tungstate crystals, which provide coverage in pseudorapidity $\abs{\eta} < 1.48$ in a barrel region (EB) and $1.48 < \abs{\eta} < 3.0$ in two endcap regions (EE).
Preshower detectors consisting of two planes of silicon sensors interleaved with a total of 3 radiation lengths of lead are located in front of each EE detector.
Forward calorimeters extend the pseudorapidity coverage provided by the barrel and endcap detectors.
Muons are measured in the pseudorapidity range $\abs{\eta} < 2.4$ by gas-ionization detectors embedded in the steel flux-return yoke outside the solenoid. 
These detectors are arranged in planes made using three technologies: drift tubes, cathode strip chambers, and resistive plate chambers.
A more detailed description of the CMS detector, together with a definition of the coordinate system used and the relevant kinematic variables, can be found in Ref.~\cite{CMS:2008xjf}.

Events of interest are selected using a two-tiered trigger system.
The first level (L1), composed of custom hardware processors, uses information from the calorimeters and muon detectors to select events at a rate of around 100\unit{kHz} within a fixed latency of about 4\mus~\cite{CMS:2020cmk}.
The second level, known as the high-level trigger, consists of a farm of processors running a version of the full event reconstruction software optimized for fast processing, and reduces the event rate to around 1\unit{kHz} before data storage~\cite{CMS:2016ngn}.

\section{Event reconstruction}\label{sec:reco}

The candidate vertex with the largest value of summed squared physics-object transverse momentum (\pt) is taken to be the primary \pp interaction vertex. 
The physics objects used in the calculation of this quantity are (i) jets, clustered using the jet finding algorithm~\cite{Cacciari:2008gp,Cacciari:2011ma}, with the tracks assigned to candidate vertices as inputs, and (ii) the associated missing \pt, taken as the negative vector sum of the \pt of those jets.

The global event reconstruction (also called particle-flow (PF) event reconstruction~\cite{CMS:2017yfk}) aims to reconstruct and identify each individual particle in an event, with an optimized combination of all subdetector information.
In this process, the identification of the particle type (photon, electron, muon, charged hadron, neutral hadron) plays an important role in the determination of the particle direction and energy.
Photons are identified as ECAL energy clusters not linked to the extrapolation of any charged particle trajectory to the ECAL.
Electrons are identified as a primary charged particle track and potentially multiple ECAL energy clusters, which correspond to the extrapolation of this track to the ECAL and to possible bremsstrahlung photons emitted along the way through the tracker material.
Muons are identified as tracks in the central tracker consistent with either a track or several hits in the muon system, and associated with calorimeter deposits compatible with the muon hypothesis.
Charged hadrons are identified as charged particle tracks that are neither identified as electrons nor as muons.
Finally, neutral hadrons are identified as HCAL energy clusters not linked to any charged hadron trajectory, or as a combined ECAL and HCAL energy excess with respect to the expected charged hadron energy deposit.

For this analysis, the detector performance for photons, electrons, and muons is critical, because the energy and momentum resolutions for these particles determine the resolution of the Higgs boson signal peak in the \mllg distribution.
The energy of photons is obtained from the ECAL measurement. In the EB, for photons that have energies in the range of tens of \GeVns, an energy resolution of about 1\% is achieved for unconverted or late-converting photons, \ie, photons converting near the inner face of the ECAL.
The energy resolution of the remaining barrel photons, which convert as they traverse the tracker material, is about 1.3\% up to $\abs{\eta} = 1$, rising to about 2.5\% at $\abs{\eta} = 1.4$.
In the EE, the energy resolution for unconverted or late-converting photons is about 2.5\%, while the remaining endcap photons have a resolution between 3 and 4\%~\cite{CMS:2015myp}.
The energy of electrons is determined from a combination of the track momentum at the main interaction vertex, the corresponding ECAL cluster energy, and the energy sum of all bremsstrahlung photons attached to the track.
The measured energy resolution for electrons produced in \PZ boson decays in \pp collision data ranges from 2--5\%, depending on electron pseudorapidity and energy loss through bremsstrahlung in the detector material~\cite{CMS:2020uim}.
The momentum of muons is obtained from the corresponding track momentum. Matching muons to tracks measured in the silicon tracker results in a \pt resolution, for muons with \pt up to 100\GeV, of 1\% in the barrel and 3\% in the endcaps.
The energy of charged hadrons is determined from a combination of the track momentum and the corresponding ECAL and HCAL energies, corrected for the response function of the calorimeters to hadronic showers.
Finally, the energy of neutral hadrons is obtained from the corresponding corrected ECAL and HCAL energies.

For tagging the VBF production mechanism, which produces an additional dijet system, jet reconstruction is important.
For each event, hadronic jets are clustered from PF particles using the infrared and collinear-safe anti-\kt algorithm~\cite{Cacciari:2008gp, Cacciari:2011ma} with a distance parameter of 0.4.
Jet momentum is determined as the vectorial sum of all particle momenta in the jet, and is found from simulation to be, on average, within 5--10\% of the true momentum over the entire \pt spectrum and detector acceptance. Additional \pp interactions within the same or nearby bunch crossings (pileup) can contribute additional tracks and calorimetric energy depositions to the jet momentum.
To mitigate this effect, charged particles found to originate from pileup vertices, \ie, any vertex other than the selected primary vertex, are discarded and an offset correction is applied to correct for remaining contributions~\cite{CMS:2020ebo}, primarily from neutral pileup particles.
Jet energy corrections are derived from simulation to bring the measured response of jets to that of particle-level jets on average.
In situ measurements of the momentum balance in dijet, photon+jet, {\PZ}+jet, and multijet events are used to account for residual differences in the jet energy scale and resolution between data and simulation~\cite{CMS:2016lmd}.
The jet energy resolution typically amounts to 15--20\% at 30\GeV, 10\% at 100\GeV, and 5\% at 1\TeV~\cite{CMS:2016lmd}.
Additional selection criteria are applied to each jet to remove jets potentially dominated by anomalous contributions from various subdetector components or reconstruction failures~\cite{CMS-PAS-JME-16-003}.

\section{Data and simulated samples}
\label{sec:samples}
The data sample corresponds to a total integrated luminosity of \LumiT\fbinv and was collected over a data-taking period spanning three years: \Lumia\fbinv in 2016, \Lumib\fbinv in 2017, and \Lumic\fbinv in 2018~\cite{CMS-LUM-17-003,LUM-17-004,LUM-18-002}.
To be considered in the analysis, events must satisfy the high-level trigger requirements for at least one of the dielectron or dimuon triggers.
The dielectron trigger requires a leading (subleading) electron with $\pt > 23\,(12)\GeV$, while the dimuon trigger requires a muon with $\pt > 17\,(8)\GeV$.
The efficiencies of these dilepton triggers, which depend on both the lepton \pt and $\eta$, are measured to be in the ranges 86--97 and 93--95\% for the electron and muon channels, respectively.

Signal samples for \ggH, VBF, \VH, and \ttH production, with \hzg and \Ztoll ($\Pell = \Pe$, \PGm, or \PGt), are generated at next-to-leading order (NLO) using \POWHEG v2.0~\cite{Nason:2004rx,Frixione:2007vw,Alioli:2010xd,cite:powheg1,cite:powheg2,Luisoni:2013cuh,Hartanto:2015uka}.
Samples are produced for \hmass of 120, 125, and 130\GeV.
The SM Higgs boson production cross sections and branching fractions recommended by the LHC Higgs Working Group~\cite{LHC-YR4,Denner:2011mq,Djouadi:1997yw,Butterworth:2010ym} are considered for each mass point.
The SM value of \brzg is calculated at LO in QCD. The effect of higher-order QCD corrections has been studied~\cite{Spira:1991tj,Bonciani:2015eua,Gehrmann:2015dua}, and found to be small.

The dominant backgrounds, $\Zgstar(\to\lplm)$+\PGg and $\Zgstar(\to\lplm)$+jets, are generated at NLO using the \MGvATNLO v2.6.0 (v2.6.1) generator~\cite{Alwall:2014hca} for 2016 (2017 and 2018) samples.
Events arising from \ttbar production~\cite{Frixione:2007nw} are a relatively minor background and are generated at NLO with \POWHEG v2.0~\cite{cite:powheg1,cite:powheg2}.
The background from vector boson scattering (VBS) production of {\Zgstar}+\PGg pairs, with the \PZ boson decaying to a pair of leptons, is simulated at leading order using the \MGvATNLO generator.
The decay \hmm is considered as a resonant background and is generated for the \ggH, VBF, \VH, and \ttH production mechanisms. The SM predicted value of the \hmm branching fraction~\cite{LHC-YR4} is assumed.
The \ggH production cross section is computed at next-to-next-to-NLO precision in QCD and at NLO in electroweak (EWK) theory~\cite{Anastasiou:2015vya,Anastasiou:2016cez}.
The cross sections for Higgs boson production in the VBF~\cite{Ciccolini:2007jr,Ciccolini:2007ec,Bolzoni:2010xr,Bolzoni:2011cu,PhysRevLett.115.082002} and \VH~\cite{Ciccolini:2003jy,BREIN2004149} mechanisms are calculated at next-to-NLO in QCD, including NLO EWK corrections, while the \ttH cross section is computed at NLO in QCD and EWK theory~\cite{Reina:2001sf,Beenakker:2001rj,Beenakker:2002nc,Dawson:2002tg,PhysRevD.68.034022,Zhang:2014gcy,Frixione:2014qaa}.

All simulated events are interfaced with \PYTHIA v8.226~(v8.230)~\cite{Sjostrand:2014zea} with the CUETP8M1~\cite{Khachatryan:2015pea} (CP5~\cite{Sirunyan:2019dfx}) underlying event tune for 2016\,(2017--2018) for the fragmentation and hadronization of partons and the internal bremsstrahlung of the leptons.
The NLO parton distribution function (PDF) set, NNPDF v3.0~\cite{nnpdf30}~(NNPDF v3.1)~\cite{nnpdf_new}, is used to produce these samples in 2016\,(2017--2018).
The response of the CMS detector is modeled using the \GEANTfour program~\cite{AGOSTINELLI2003250}.
The simulated events are reweighted to correct for differences between data and simulation in the number of additional \pp interactions, trigger efficiencies, selection efficiencies, and efficiencies of isolation requirements for photons, electrons, and muons.

\section{Event selection}\label{sec:preselection}

Events are required to have at least one good primary vertex (PV) with a reconstructed longitudinal position within 24\cm of the geometric center of the detector and a transverse position within 2\cm of the nominal beam collision point.
Lepton candidates are required to have impact parameters with respect to the PV of less than 5\mm in the plane transverse to the beam and less than 10\mm along the beam direction.

This analysis focuses on promptly produced signal processes.
To reduce the contributions from photons or leptons arising from hadron decays within jets, isolation requirements are imposed.
For each photon and lepton candidate, a set of isolation variables is defined.
The quantity \ptchg is the scalar sum of the \pt of charged hadrons originating from the PV, and \ptneut and \ptgam are the scalar sums of the \pt of neutral hadrons and photons, respectively.
The sums are over all PF candidates within a cone of radius $\DR = \sqrt{\smash[b]{(\Delta\phi)^2 + (\Delta\eta)^2}} = 0.3$ around the photon or lepton direction at the PV.

Photons are selected with an MVA discriminant that uses, as inputs, the isolation variables \ptchg, \ptneut, and \ptgam; the ratio of the HCAL energy to the sum of the ECAL and HCAL energies associated with the cluster; and the transverse width of the electromagnetic shower.
The imperfect MC simulation modeling of the input variables is corrected to match the data using a chained quantile regression method~\cite{2012arXiv1211.6581S} based on studies of \Ztoee events.
In this method, a set of boosted decision tree (BDT) discriminants is trained to predict the cumulative distribution function for a given input.
Its prediction is conditional upon the three kinematic variables (\pt, $\eta$, $\phi$) and the global event energy density~\cite{CMS:2018piu}, which are the input variables to the BDTs.
The corrections are then applied to the simulated photons such that the cumulative distribution function of each simulated variable matches that observed in data.
A conversion-safe electron veto~\cite{CMS:2015myp} is applied to reduce the contribution from electrons that are misidentified as photons.
This veto suppresses events that have a charged particle track with a hit in the inner layer of the pixel detector that points to the photon cluster in the ECAL, unless that track is matched to a conversion vertex. This typically rejects 95\,(80)\% of the misidentified electrons in the barrel (endcaps).
Photons are required to lie in the geometrical region $\abs{\eta} < 2.5$.
The efficiency of the photon identification is measured from \Ztoee events using the ``tag-and-probe" technique~\cite{cite:tagandprobe}.
The efficiency is measured as a function of the probe candidate \pt, and covers the \pt range of photons selected in the analysis. It is measured separately for barrel and endcap candidates, resulting in efficiencies that vary with the photon \pt from 76-–90 and 72–-90\%, respectively, which include the electron veto~\cite{CMS:2015myp} inefficiencies measured with $\Ztomm\PGg$ events, where the photon is produced by final-state radiation (FSR).

Electrons are selected using an MVA discriminant that includes observables sensitive to the shape of the electromagnetic shower in the ECAL, the geometrical and momentum-energy matching between the electron trajectory and the energy of the associated cluster in the ECAL, the presence of bremsstrahlung along the electron trajectory, isolation, and variables that discriminate against electrons originating from photon conversions~\cite{bib:htozz2016}.
The electron MVA discriminant includes the isolation sums described above (\ptchg, \ptneut, and \ptgam).
Electron candidates must satisfy $\abs{\eta} <2.5$.
The optimized electron selection criteria give an efficiency of approximately 85--93\,(81--92)\% in the barrel (endcaps), depending on \pt, for electrons from \PW or \PZ bosons.

Muons are selected from the reconstructed muon track candidates by applying minimal requirements on the track in both the muon system and inner tracker system and by taking into account compatibility with small energy deposits in the calorimeters.
A muon isolation requirement is used to veto potential muon candidates that are produced in the decays of heavy quarks.
We define the muon relative isolation
\begin{linenomath*}
\begin{equation}
\label{eqn:pfiso}
\mathcal{I}^{\PGm} \equiv \Big[ \ptchg +
                                 \max\big(0, \ptneut
                                 +
                                 \ptgam
                                 - \ptpu \big) \Big]
                                 / \pt^{\PGm}
\end{equation}
\end{linenomath*}
and require $\mathcal{I}^{\PGm} < 0.35$.
Since the isolation variable is particularly sensitive to energy deposits from pileup interactions, a \ptpu contribution is subtracted, defined as $\ptpu\equiv0.5\sum_{i} \pt^{\text{PU},i}$, where $i$ runs over the momenta of the charged hadron PF candidates not originating from the PV, and the factor of 0.5 corrects for the different fraction of charged and neutral particles in the cone~\cite{CMS:2020ebo}.
The combined identification and isolation efficiency for single muons is measured using \Ztomm decays and is found to be 87--98\% in the barrel region and 88--98\% in the endcaps, depending on the muon \pt.
We accept muons with $\abs{\eta}<2.4$ \cite{bib:htozz2016}.
To suppress backgrounds in which muons are produced in the decays of hadrons and electrons from photon conversions, we require each muon track to have a three-dimensional impact parameter with respect to the PV that is less than four times its uncertainty.

An FSR recovery procedure is performed for the selected muons, following a similar approach to that used in Ref.~\cite{bib:htozz2016}.
An FSR photon is identified and associated to its radiating muon based on the following criteria.
The photon must satisfy $\pt>2\GeV$, $\abs{\eta} < 2.4$, $\DRgmRatio < 0.012 \GeV^{-2}$, $\DRgm < 0.4$, and relative isolation smaller than 1.8, where $\pt^{\text{PU}}$ is excluded from the isolation calculation.
If multiple FSR photons are associated to one muon, the photon with the smallest value of \DRgmRatio is selected.
The FSR recovery procedure improves the \mllg resolution by 1\% in the muon channel.

A kinematic fit procedure is used to improve the dilepton mass and \mllg resolutions, following a similar approach to that used in Ref.~\cite{bib:htozz2016}.
A maximum likelihood fit is performed, taking into account the true \PZ boson line shape, obtained from \hzg simulation, the \pt of each lepton,
and the \pt resolution of each lepton.
The outputs of this fit are the corrected \pt values for each lepton.
The corrected \pt values are used to recalculate the dilepton mass and \mllg.
The improvement in \mllg resolution varies with data-taking year and is between 20--27\% in the electron channel and 10--12\% in the muon channel. The effect of the kinematic fit is larger for the electron channel because of the poorer momentum resolution for electrons compared to muons.

The jets used in dijet-tagged event categories, discussed in Section~\ref{sec:class}, are required to have $\pt > 30\GeV$ and $\abs{\eta}<4.7$ and to be separated by at least 0.4 in $\DR$ from leptons and photons passing the selection requirements described above.

Events are required to contain a photon and at least two same-flavor, opposite-sign leptons ($\Pell = \Pe$ or \PGm) with $\mll>50\GeV$.
The latter requirement, although relatively loose, is sufficient to suppress backgrounds that do not contain \PZ boson decays while retaining high signal efficiency.
The particles used to reconstruct the \zg candidate system are required to have $\pt > 25\,(15)\GeV$ for the leading (subleading) electron, $\pt > 20\,(10)\GeV$ for the leading (subleading) muon, and $\pt > 15\GeV$ for the photon.
In events with multiple dilepton pairs, the pair with mass closest to the nominal \PZ boson mass~\cite{PDG2020} is selected. In signal simulation, the correct leptons are identified in more than 99\% of events.
Additional electrons (muons) with \pt greater than 7\,(5)\GeV are also used for categorization, as described in the next section.

The invariant mass of the \llg system is required to be in the range $105 < \mllg < 170\GeV$, which provides a broad range around the Higgs boson mass in which to perform the fit.
Events are required to have a photon satisfying $\pt^{\PGg} / \mllg > 0.14$, which suppresses the {\Zgstar}+jets background
without significantly reducing signal efficiency, with minimal bias in the \mllg spectrum.
Each lepton is required to have $\DR>0.4$ with respect to the photon to reject events with FSR.
To further reject FSR from {\Zgstar}+\PGg processes, we require $\mllg + \mll> 185$\GeV.

\section{Event categorization}
\label{sec:class}

To maximize the sensitivity of the analysis to Higgs boson signals arising from different production mechanisms, each with its own final-state properties, we divide the event sample into mutually exclusive categories.
The signal candidates from the \VH and \ttH production mechanisms are targeted using a lepton-tagged category, in which at least one electron or muon is present beyond those used to reconstruct the \zg system.
The signal candidates from the VBF production mechanism are targeted by identifying events that have an additional dijet system.
A BDT classifier (referred to as the VBF BDT) uses the properties of this dijet system to divide such events into a set of dijet categories.
The VBF BDT discriminant value, transformed such that the VBF signal distribution is uniform, is denoted by \DVBF.
The signal candidates from the \ggH production mechanism are targeted with events that do not fall within the lepton-tagged or dijet categories.
A BDT classifier (referred to as the kinematic BDT), trained on a set of kinematic variables, is used to further discriminate between signal and background events, defining a set of untagged categories.
The kinematic BDT discriminant value, transformed such that the total signal distribution is uniform, is denoted by \Dkin.

The procedure used for event categorization is described below.

\begin{figure}[tbp]
\centering
\includegraphics[width=0.47\textwidth]{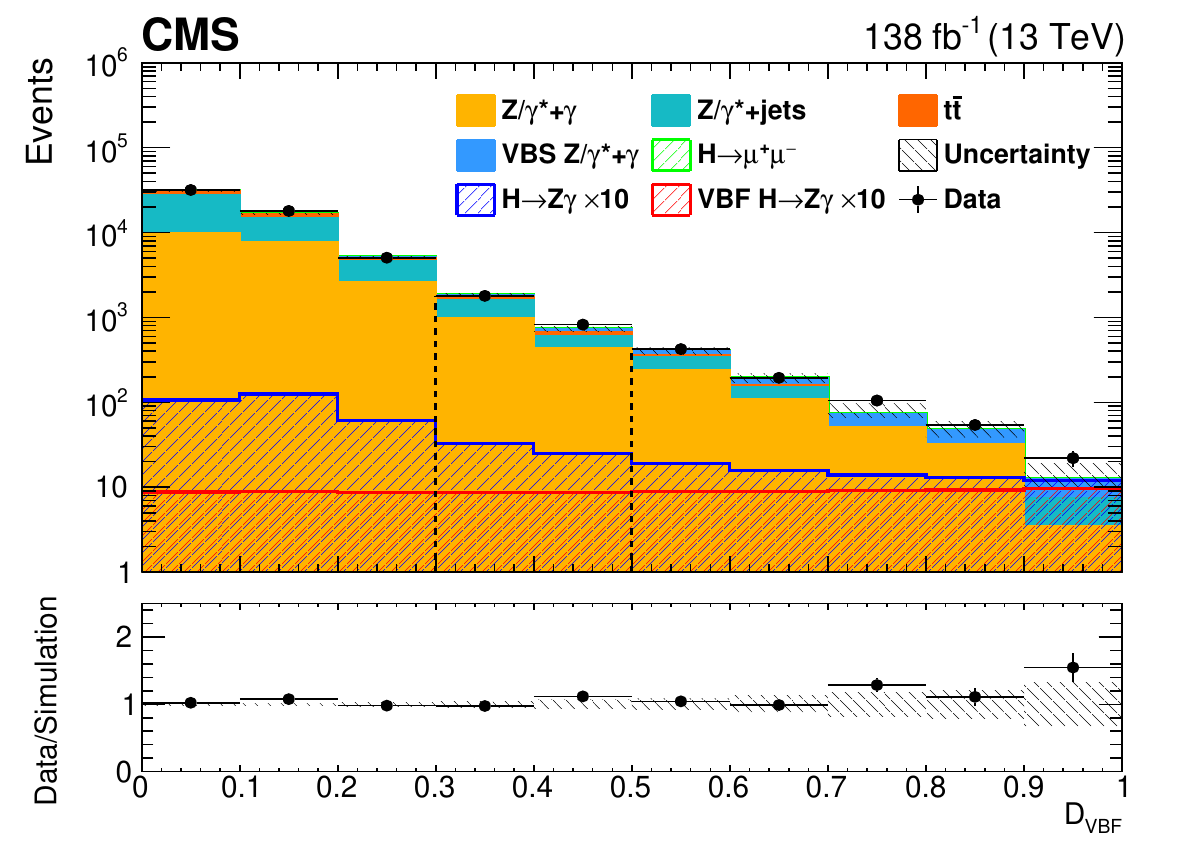}
\includegraphics[width=0.47\textwidth]{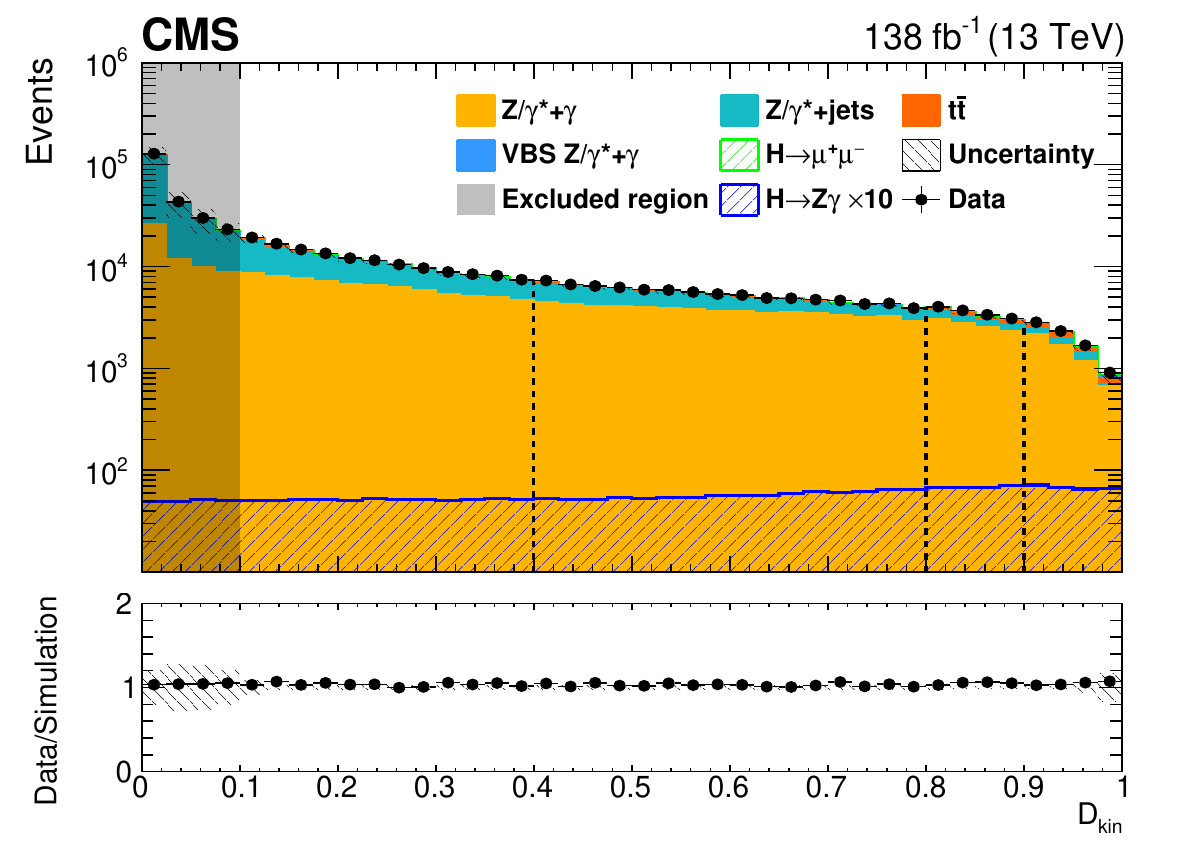}
 \caption{The \DVBF (left) and \Dkin (right) distributions for signal, simulated background, and data.
    The \DVBF distribution includes only dijet-tagged events, and the \Dkin distribution includes only untagged events. 
    The sum of contributions from all signal production mechanisms is shown by the blue line, while the contribution from only the VBF mechanism is shown by the red line. Both contributions are scaled by a factor of 10.
    The uncertainty band incorporates all statistical and systematic uncertainties in the expected background.
    The dashed lines indicate the boundaries for the dijet and untagged categories.
    The gray shaded region in the \Dkin distribution is excluded from the analysis.}\label{fig:BDT}
\end{figure}

\begin{enumerate}
  \item Events with at least one additional electron (muon) with $\pt>7\,(5)\GeV$ are assigned to the lepton-tagged category. These thresholds, which are lower than those used for the leptons in the \hzg decay, are chosen to maximize the signal acceptance for the associated production processes.
  \item Events not assigned to the lepton-tagged category, but which contain two jets satisfying the selection requirements described in Section~\ref{sec:preselection}, are classified as dijet events, indicative of possible VBF production.
  If multiple dijet pairs exist within an event, the two jets with highest \pt are considered.
  The subdivision of dijet events into a set of three dijet categories is described later in this section.
  A VBF BDT classifier is trained to separate VBF signal events from {\ggH}+jets and background events.
  The following variables are used in the VBF BDT training:
  (i)~the difference in pseudorapidity between the two jets;
  (ii)~the difference in azimuthal angle between the two jets;
  (iii)~the Zeppenfeld variable~\cite{Rainwater:1996ud} ($\eta_{\PGg} - (\eta_{\mathrm{j_1}}+\eta_{\mathrm{j_2}})/2$), where $\eta_{\PGg} ,\eta_{\mathrm{j_1}}$ and $\eta_\mathrm{{j_2}}$ are the pseudorapidities of the photon, leading jet, and subleading jet, respectively;
  (iv)~the ratio between the \pt of the $\zg{\mathrm{j_1}}{\mathrm{j_2}}$ system and the corresponding scalar sum of momenta
    ($\abs{\sum_{\PZ,\PGg,\mathrm{j_1},\mathrm{j_2}}{\ptvec}} / \sum_{\PZ,\PGg,\mathrm{j_1},\mathrm{j_2}}p_\mathrm{T}$);
  (v)~the difference in azimuthal angle between the dijet system and the \zg system;
  (vi)~the \pt of each jet;
  (vii)~$\pt^{t}$, defined as $\abs{\ptvec^{\zg}\times\hat{t}}$, where $\hat{t}=(\ptvec^{\PZ}-\ptvec^{\PGg})/\abs{\ptvec^{\PZ}-\ptvec^{\PGg}}$~\cite{Ackerstaffetal.1998,VESTERINEN2009432}, the \pt of the \zg system that is perpendicular to the difference of the three-momenta of the \PZ boson and the photon, a quantity that is strongly correlated with the \pt of the \llg system;
  (viii)~the $\DR$ separation between each jet and the photon, and (ix) \Dkin, described below. The distribution of \DVBF is shown in Fig.~\ref{fig:BDT} (left) for both simulated event samples and data.
  \item Events not assigned to the lepton-tagged or dijet categories are classified as untagged events.
  The subdivision of untagged events into a set of four untagged categories is described later in this section.
  A kinematic BDT classifier is trained to distinguish signal events from background events based on the kinematics of the leptons and photon in the \zg candidate system, as well as on the measured properties of these physics objects. Signal events from all production processes are included in the training.
  The following variables are used in the kinematic BDT training:
  (i)~the pseudorapidity of each lepton and the photon;
  (ii)~the $\DR$ separation between each lepton and the photon;
  (iii)~the \pt to mass ratio of the \llg system;
  (iv)~the production angle of the \PZ boson in the Higgs boson center-of-mass frame~\cite{HZg_angle1,HZg_angle2};
  (v)~the polar and azimuthal decay angles of the leptons in the \PZ boson center-of-mass frame~\cite{HZg_angle1,HZg_angle2};
  (vi)~the photon MVA discriminant score; and (vii) the photon energy resolution.
  The distribution of \Dkin is shown in Fig.~\ref{fig:BDT} (right) for both simulated samples and data.
\end{enumerate}

The subdivision of dijet and untagged events into categories is based on the VBF BDT and kinematic BDT discriminants.
Category boundaries are defined as mutually exclusive regions of \DVBF and \Dkin.
The locations of the boundaries defining the categories are optimized by iterating over all possible combinations of boundaries using $\sum_{i=1}^{n} {S_i}^2/B_i$ as a figure-of-merit.
The variables $S_i$ and $B_i$ represent the number of expected signal and background events with $120 < \mllg < 130\GeV$ in the $i$th category, and $n$ is the total number of categories.
We consider categories with boundaries corresponding to signal efficiencies between 0--100\% in 10\% increments.
The optimization procedure results in three dijet categories for the VBF BDT and four untagged categories for the kinematic BDT.
The lowest \Dkin boundary corresponds to the 10\% point in integrated signal efficiency, and events below the 10\% point are excluded from the analysis, as they make a negligible contribution to the total signal sensitivity.

The full categorization and optimization procedure results in the following eight mutually exclusive categories: one lepton-tagged category, three dijet categories, and four untagged categories.
The category definitions are summarized in Table~\ref{tab:category}.

\begin{table}[tbp]
  \centering
  \topcaption{Summary of the category definitions. The lepton-tagged category requires at least one additional electron or muon. Dijet categories are defined by regions of \DVBF and untagged categories are defined by regions of \Dkin.}
  \cmsTable{
  \begin{tabular}{c@{\hskip 0.3in}ccc@{\hskip 0.3in}cccc}
    \hline
    Lepton                              & Dijet 1 & Dijet 2 & Dijet 3& Untagged 1 & Untagged 2 & Untagged 3 & Untagged 4 \\
    \hline
    \multirow{2}{*}{$\geq$ 1 $\Pe,\mu$} &\multicolumn{3}{c}{\DVBF selection}&\multicolumn{4}{c}{\Dkin selection} \\
                                        &0.5--1.0&0.3--0.5&0.0--0.3&0.9--1.0&0.8--0.9&0.4--0.8&0.1--0.4\\
    \hline
  \end{tabular}}\label{tab:category}
\end{table}

Table~\ref{tab:yield} lists the event categories used in the analysis, along with the expected event yields for an $\hmass=\mH\GeV$ signal arising from \ggH, VBF, \VH, and \ttH production, as well as the resonant background contribution from FSR from \hmm, which is 3--8\% of the \hzg yield, depending on category.
Event yields from other Higgs boson backgrounds such as $\PH\to\PGtp\PGtm$ and \hgg are estimated to be below the 1\% level relative to the \hzg yield and are neglected.
The dominant contribution to the signal yield is generally from \ggH production, except in the lepton-tagged category, in which \VH and \ttH events dominate, and in the dijet 1 category, in which VBF events dominate.
The categorization procedure increases the sensitivity of the analysis by 24\% with respect to an inclusive event selection.
The product of signal acceptance and efficiency for $\pp\to\hzg\to\llg$ for $m_{\PH} =\mH\GeV$ is 23\,(29)\% in the electron (muon) channel.

  \begin{table}[tbp]
    \centering
    \topcaption{Yields and approximate significance ($S/\sqrt{B}$) for each category, where $S$ and $B$ are the expected number of signal and background events in the narrowest \mllg interval containing 95\% of the expected signal distribution.
    Also shown is the \mllg resolution, computed using the narrowest interval containing 68\% of the expected signal distribution.
     }
      \cmsTable{
  \begin{tabular}{c @{\hskip 0.2in}ccccccccc}
    \hline
    \LumiT\fbinv   & Lepton   &  & Dijet 1 & Dijet 2 & Dijet 3& Untagged 1 & Untagged 2 & Untagged 3 & Untagged 4\\\hline
    SM signal    &&        &           &           &           &        &        &        &       \\
    yield        &&        &           &           &           &        &        &        &       \\
    \multirow{2}{*}{\ggH}  & \multirow{2}{*}{0.51} & \epem & 1.10 & 1.62 & 9.44 & 6.89 & 7.35 & 29.8 & 22.5 \\
                                  &                       & \mpmm & 1.41 & 2.05 & 12.1 & 8.52 & 9.17 & 38.0 & 29.0 \\ [\cmsTabSkip]
    \multirow{2}{*}{VBF}          & \multirow{2}{*}{0.09} & \epem & 1.94 & 0.76 & 1.13 & 0.71 & 0.35 & 0.92 & 0.51 \\
                                  &                       & \mpmm & 2.40 & 0.97 & 1.43 & 0.89 & 0.43 & 1.18 & 0.65 \\ [\cmsTabSkip]
    \multirow{2}{*}{$\VH+\ttH$} & \multirow{2}{*}{1.84} & \epem & 0.04 & 0.13 & 1.89 & 0.31 & 0.17 & 0.45 & 0.27 \\
                                        &                       & \mpmm & 0.05 & 0.16 & 2.36 & 0.39 & 0.21 & 0.57 & 0.33 \\ [\cmsTabSkip]
    SM resonant &&&&&&&&&\\
    background &&&&&&&&&\\
  \hmm     & 0.14& \mpmm &0.27 & 0.27 & 0.43& 0.62& 0.49 & 2.02& 1.78\\ \hline 

  Mass resolution & \multirow{2}{*}{2.12} & \epem & 1.91 & 2.06 & 2.15 & 1.80 & 1.97 & 2.12 & 2.33 \\
  (\GeVns)          &                     & \mpmm & 1.52 & 1.61 & 1.72 & 1.37 & 1.42 & 1.62 & 1.83 \\ [\cmsTabSkip]
  Data yield            & 1485   && 168    & 589    & 11596& 1485      & 1541      & 2559      & 17608   \\ [\cmsTabSkip]
  $S/\sqrt{B}$    & 0.06  && 0.54  & 0.24  & 0.26& 0.45  & 0.35  & 0.53  & 0.30  \\
  \hline
  \end{tabular}
    }
  \label{tab:yield}
  \end{table}

\section{Statistical procedure}
\label{sec:modeling}

\begin{figure*}[tbp]
  \centering
  \includegraphics[width=0.45\textwidth]{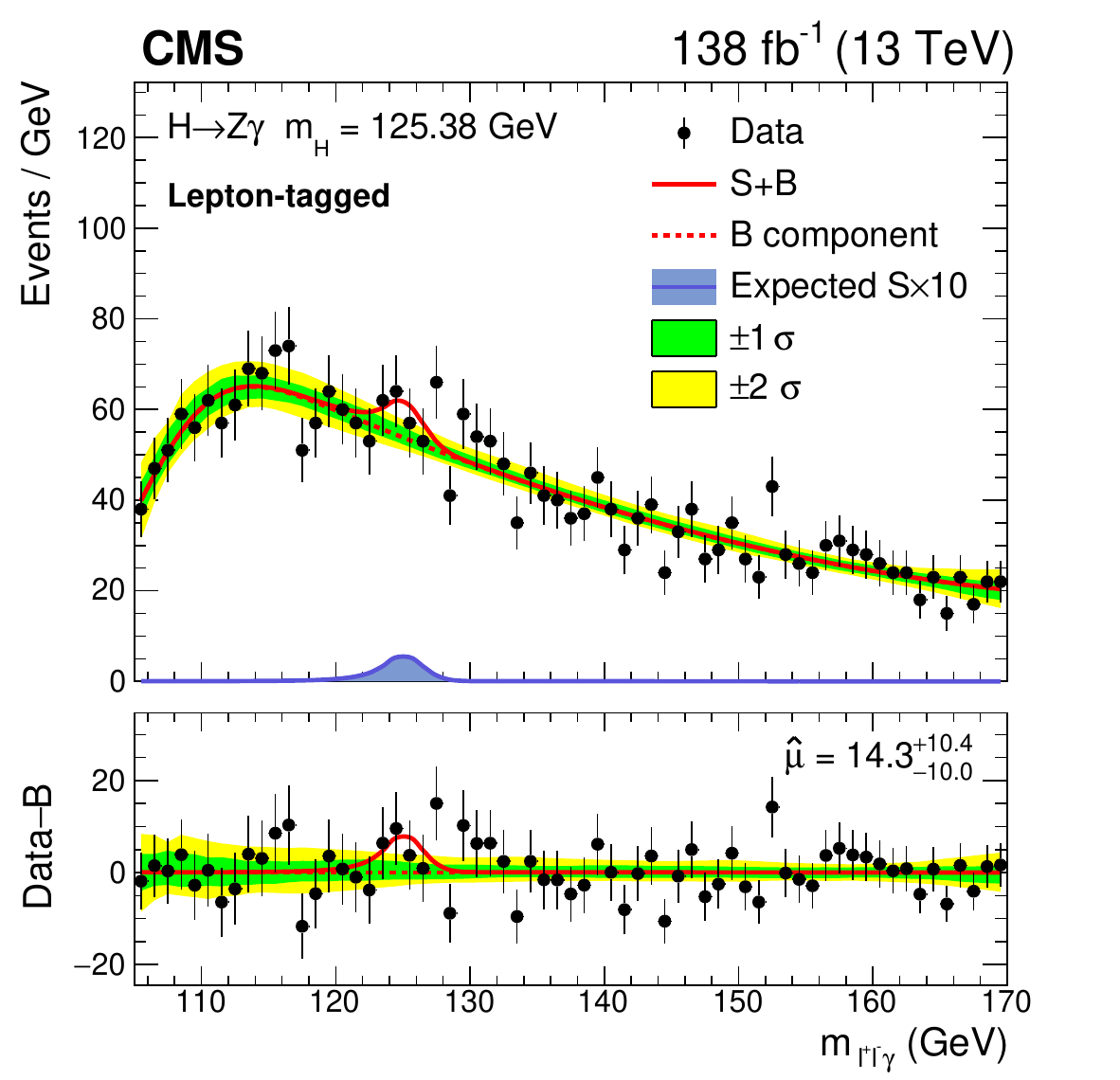}
  \includegraphics[width=0.45\textwidth]{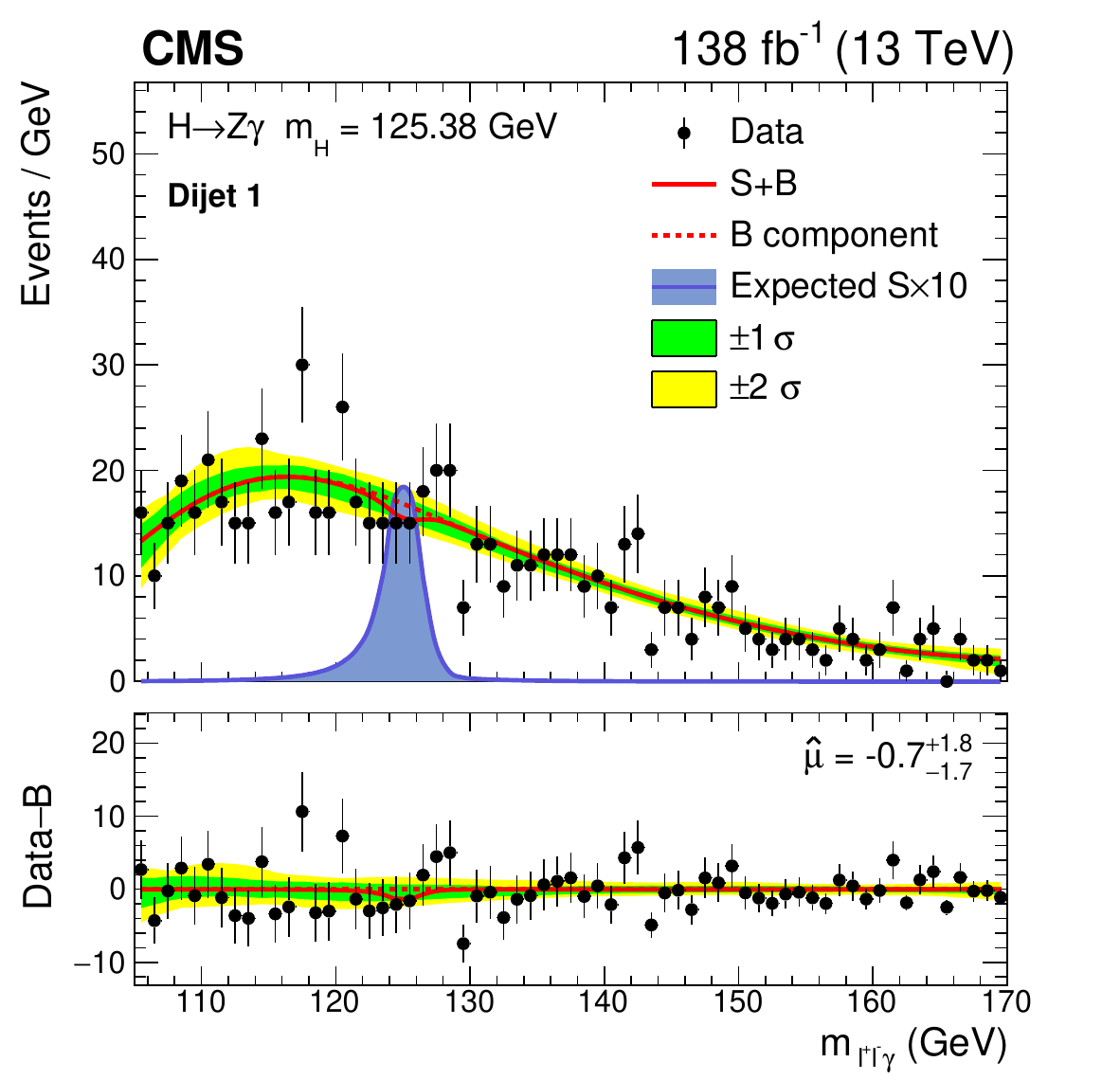}\\
  \includegraphics[width=0.45\textwidth]{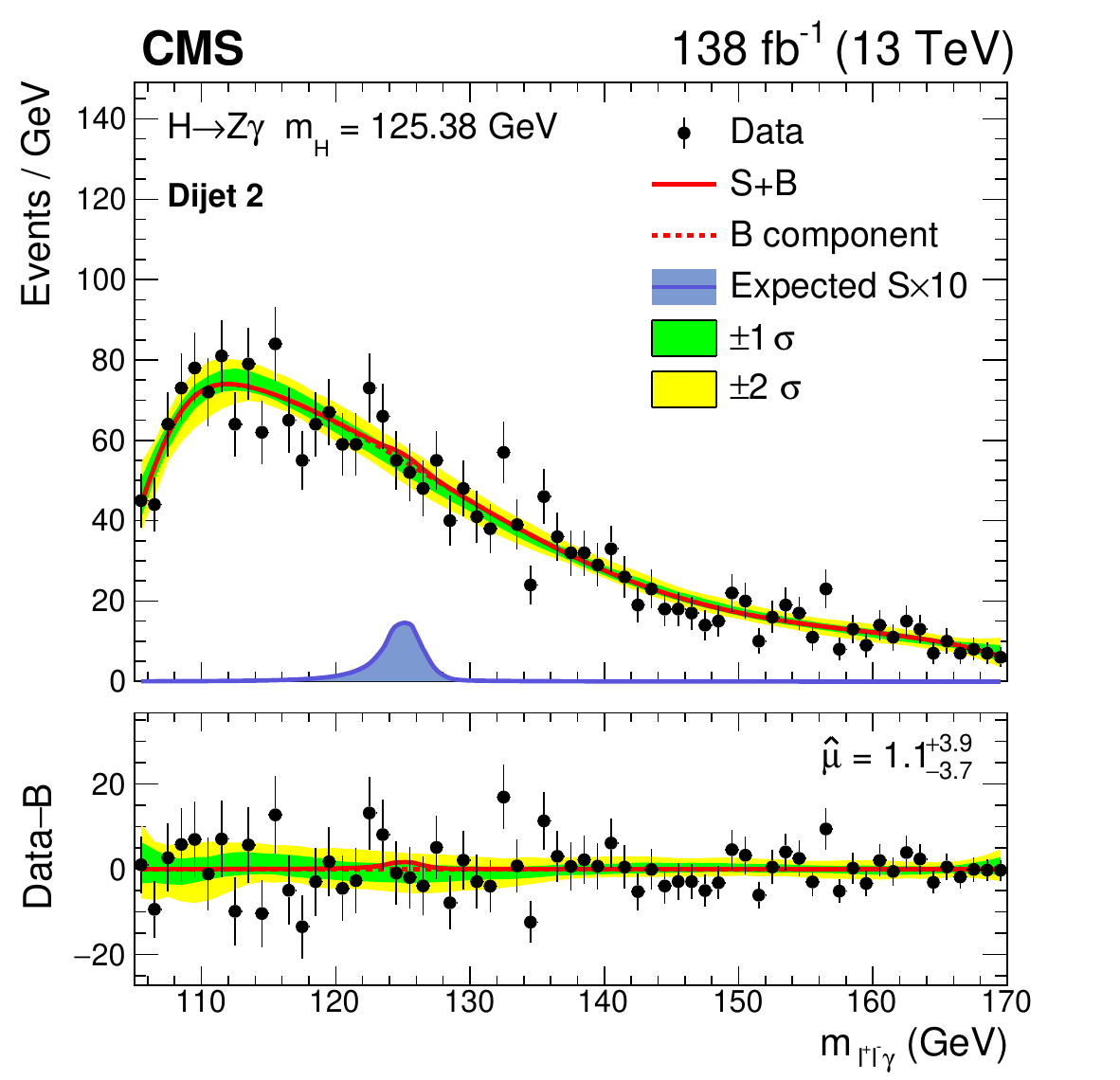}
  \includegraphics[width=0.45\textwidth]{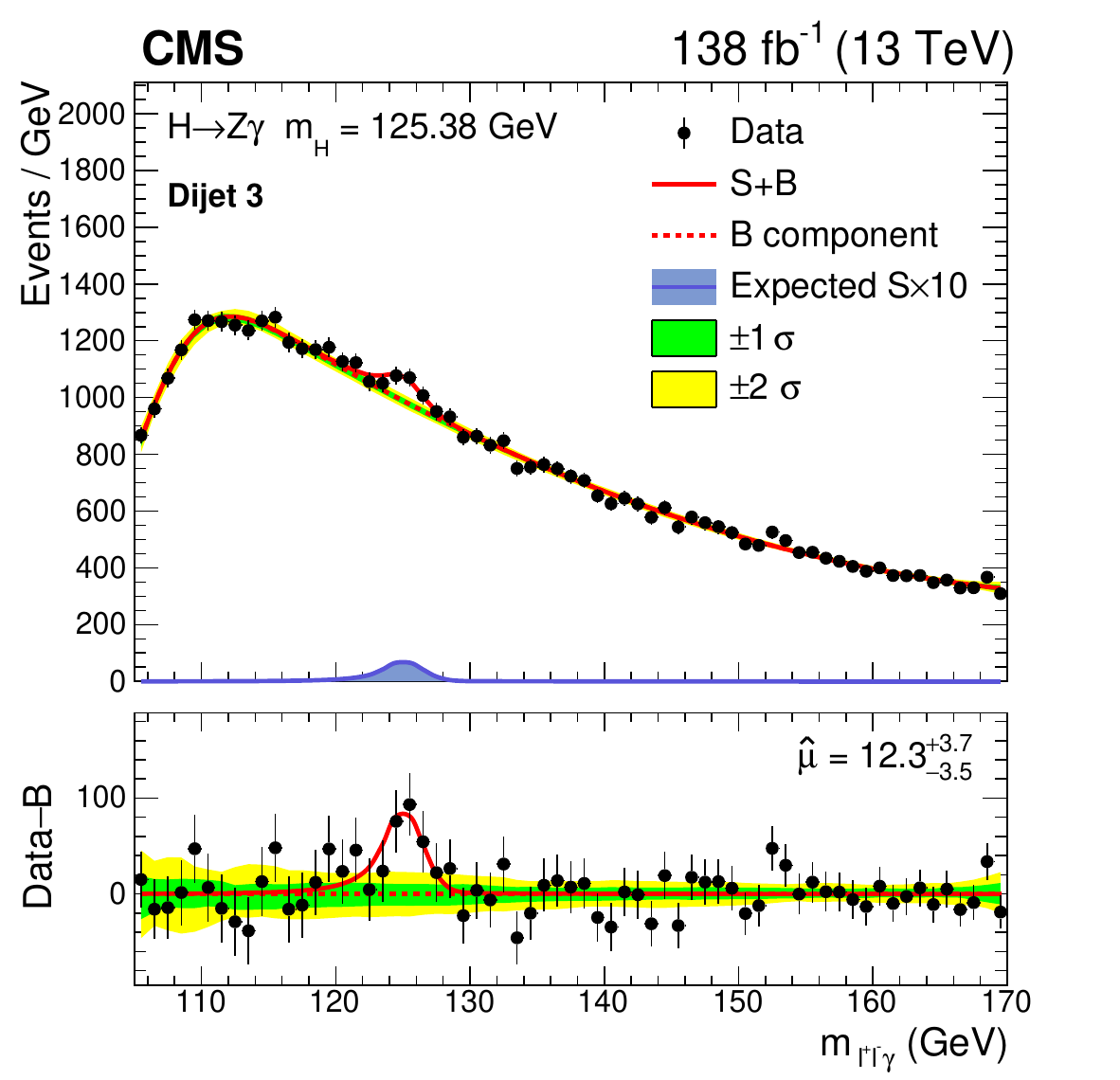}\\
   \caption{Fits to the \mllg data distribution in the lepton-tagged (upper left), dijet 1 (upper right), dijet 2 (lower left), and dijet 3 (lower right) categories.
  In the upper panel, the red solid line shows the result of a signal-plus-background fit to the given category.
  The red dashed line shows the background component of the fit.
  The green and yellow bands represent the 68 and 95\% \CL uncertainties in the fit.
  Also plotted is the expected SM signal, scaled by a factor of 10.
  In the lower panel, the data minus the background component of the fit is shown.}\label{fig:3}
  \end{figure*}

\begin{figure*}[tbp]
  \centering
  \includegraphics[width=0.45\textwidth]{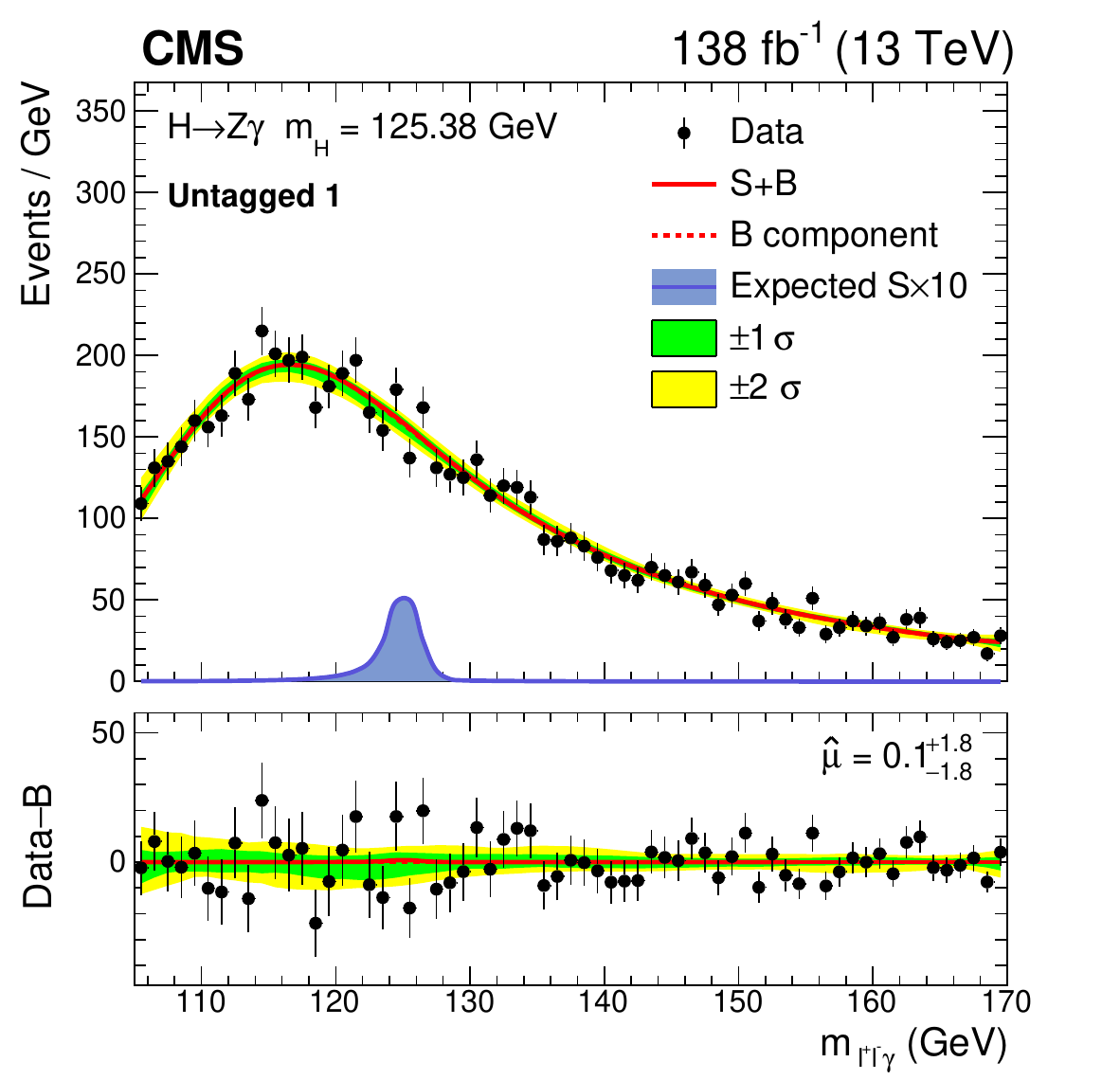}
  \includegraphics[width=0.45\textwidth]{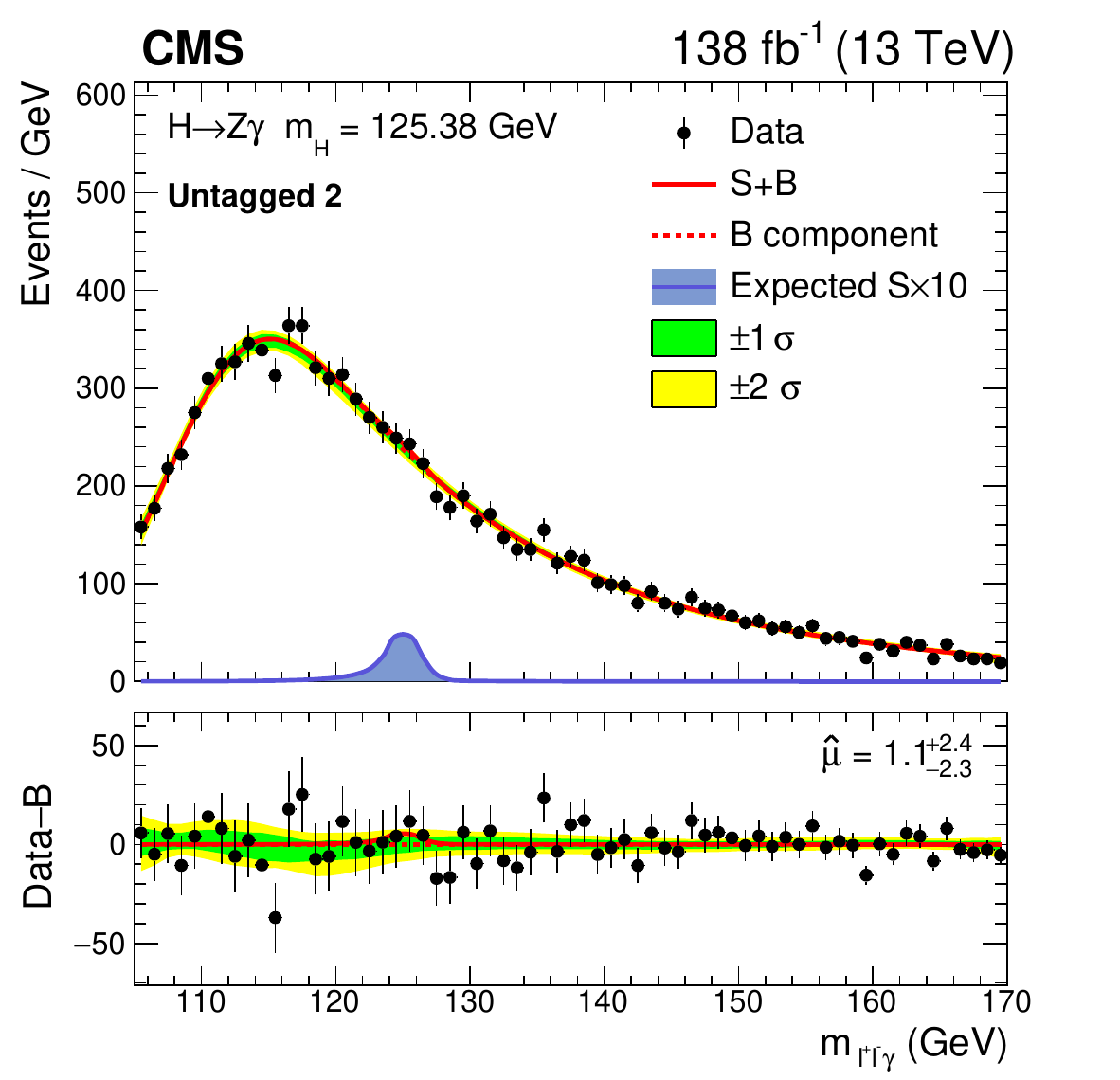}\\
  \includegraphics[width=0.45\textwidth]{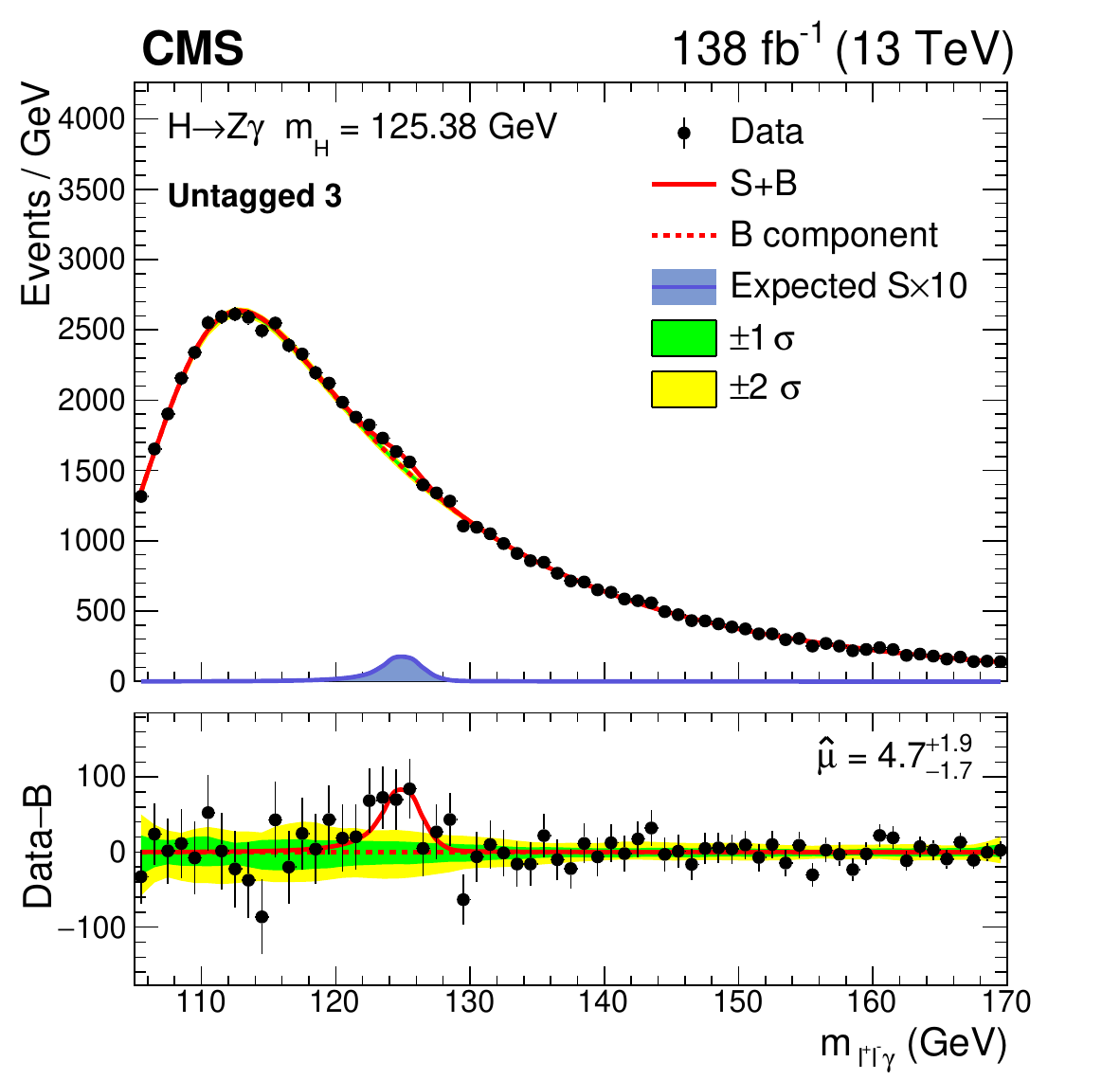}
  \includegraphics[width=0.45\textwidth]{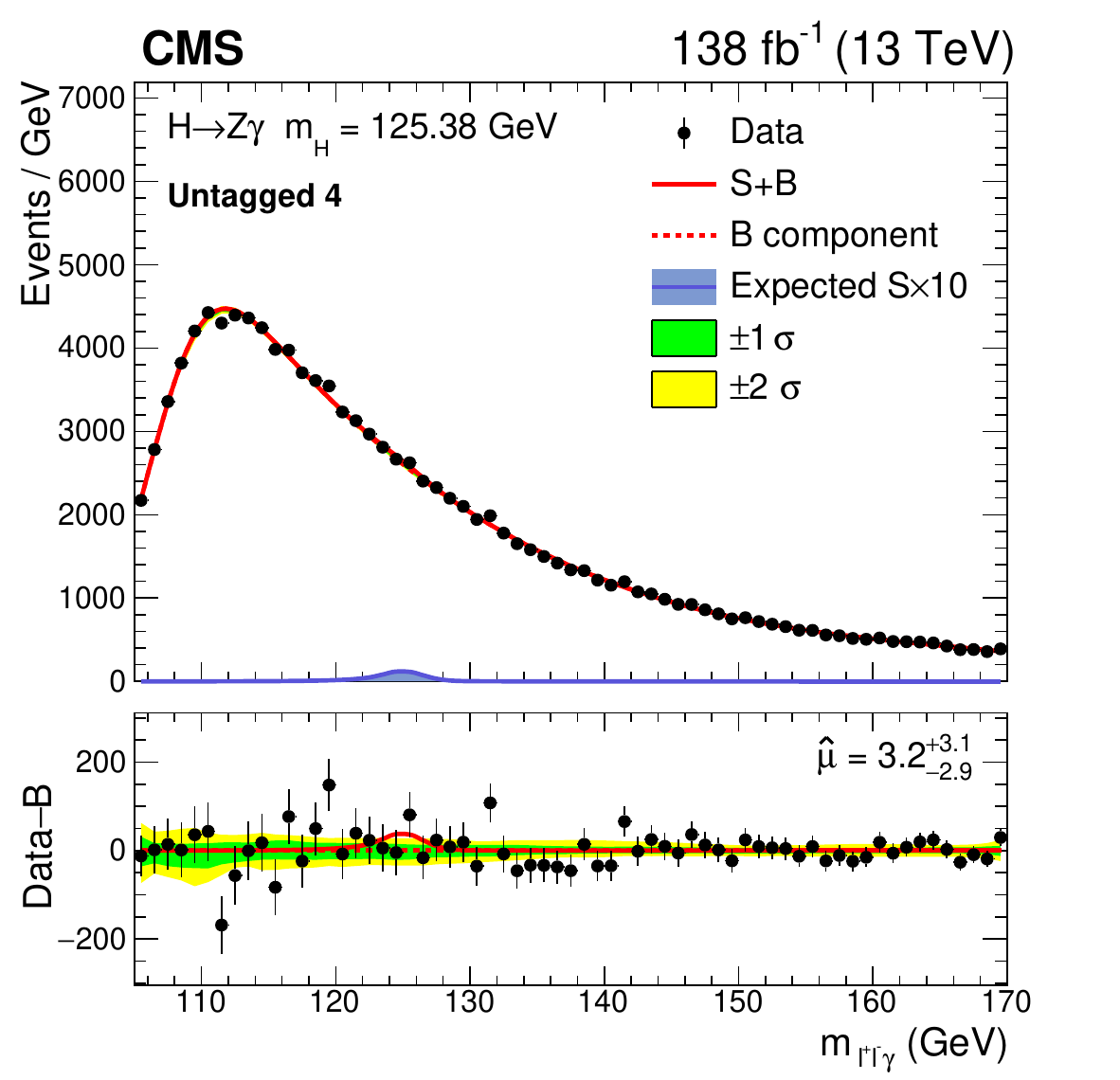}
   \caption{Fits to the \mllg data distribution in the untagged 1 (upper left), untagged 2 (upper right), untagged 3 (lower left), and untagged 4 (lower right) categories.
  In the upper panel, the red solid line shows the result of a signal-plus-background fit to the given category.
  The red dashed line shows the background component of the fit.
  The green and yellow bands represent the 68 and 95\% \CL uncertainties in the fit.
  Also plotted is the expected SM signal, scaled by a factor of 10.
  In the lower panel, the data minus the background component of the fit is shown.}\label{fig:4}
  \end{figure*}

The signal search is performed using a simultaneous fit to the \mllg distribution in the eight event categories described in Section~\ref{sec:class}.
Figures~\ref{fig:3} and \ref{fig:4} show the \mllg distributions of the data events in each category.
The expected SM \hzg distributions, scaled by a factor of 10, are also shown.
The fit uses a binned maximum likelihood method in the range $105 < \mllg < 170\GeV$.
In each category, a likelihood function is defined using analytic models of signal and background events, along with nuisance parameters for systematic uncertainties.
The combined likelihood function is the product of the likelihood functions in each category.
The parameter of interest in the maximum likelihood fit is the signal strength $\mu$, defined as the product of the cross section and the branching fraction [$\sigxbr$], relative to the SM prediction. The fit results shown in Figs.~\ref{fig:3} and \ref{fig:4} are discussed further in Section~\ref{sec:results}.

The signal model is defined as the sum of Crystal Ball~\cite{CB-Oreglia} and Gaussian functions.
The signal shape parameters are determined by fitting this model to simulated signal events in each category.
To account for differences in mass resolution, these fits are performed separately for the event samples used to model each data-taking year, as well as for muon and electron channel events.
This results in six signal models that are summed to give the total signal expectation in a given category.
Table~\ref{tab:yield} gives these mass resolutions for \hzg, summed over the three years, as obtained from simulation.
The mass resolutions range from 1.4--2.3\GeV, depending on the category.
Separate sets of parameter values are found by fitting simulated events with \hmass of 120, 125, and 130\GeV.
Using linear interpolation, parameter values are also determined at 1\GeV intervals in \hmass from 120--130\GeV, as well as at \mH\GeV.
In the fit to data, the mean and resolution parameters are allowed to vary subject to constraints from several systematic uncertainties, described in Section~\ref{sec:systematics}, while the remaining parameters are held fixed.
The resonant background contribution from \hmm is also modeled with the sum of Crystal Ball and Gaussian functions, using an analogous procedure.

The background model in each category is obtained from the data using the discrete profiling method~\cite{Dauncey:2014xga}.
This technique accounts for the systematic uncertainty associated with choosing an analytic functional form to fit the background.
The background function is chosen from a set of candidate functions via a discrete nuisance parameter in the fit.
These functions are derived from the data in each category, with muon and electron events from all data-taking years combined.
As shown in Figs.~\ref{fig:3} and \ref{fig:4}, the \mllg spectrum consists of a turn-on peak around 110--115\GeV, driven by the photon \pt selection, and a monotonically falling spectrum in the higher \mllg region.
These features are modeled by the convolution of a Gaussian function, which is used to describe the lower-mass (turn-on) portion of the spectrum, with a step function that is multiplied by one of several functions, which are used to describe the higher-mass (tail) portion of the spectrum.
The complete function has the general form:
\begin{linenomath*}
\begin{equation}
    \mathcal{F}(\mllg; \mu_{\mathrm{G}}, \sigma_{\mathrm{G}}, s, \vec{\alpha}) = \int_{-\infty}^{\infty}\mathcal{N}(\mllg-t;\mu_{\mathrm{G}},\sigma_{\mathrm{G}})\Theta(t; s)f(t; \vec{\alpha})\rd t,
\end{equation}
\end{linenomath*}
where $t$ is the integration variable for the convolution, $\mathcal{N}(\mllg-t;\mu_{\mathrm{G}},\sigma_{\mathrm{G}})$ is the Gaussian function with mean $\mu_{\mathrm{G}}=0$ and standard deviation $\sigma_{\mathrm{G}}$, $\Theta(t; s)$ is the Heaviside step function with step location $s$, and $f(t; \vec{\alpha})$ is the falling spectrum function with shape parameters $\vec{\alpha}$.
The falling spectrum function families considered include exponential functions, power law functions, Laurent series, and Bernstein polynomials.
Functions from each family are selected based on a chi-squared goodness-of-fit criterion ($p\text{-value} > 0.01$) as well as an $\mathcal{F}$-test~\cite{Fisher:1922saa}, which determines the highest order function to be used.
A penalty term is added to the final likelihood to take into account the number of parameters in each function, ensuring that higher-order functions will not be preferred a priori.
The set of profiled background functions in each category is checked to ensure that any bias introduced into the fit results is small and that the associated \CL intervals have the appropriate frequentist coverage.
For each function, pseudo-data sets are generated under a fixed signal strength hypothesis.
A signal-plus-background fit is performed on each pseudo-data set, with the choice of background function profiled.
The average bias, expressed as a fraction of the signal strength uncertainty, is typically 2--10\%, depending on the category and choice of function, and the corresponding coverage for the 68\% \CL interval is 66--69\%.

The best fit value of the signal strength, $\hat{\mu}$, is determined by maximizing the likelihood, accounting for all nuisance parameters.
The uncertainty in $\hat{\mu}$ and the observed significance are derived from the profile likelihood test statistic~\cite{cite:l3},
\begin{linenomath*}
\begin{equation}
  q(\mu) = -2\ln\Bigg(\frac{\mathcal{L}(\mu, \hat{\vec{\theta_{\mu}}})}{\mathcal{L}(\hat{\mu},\hat{\vec{\theta}})}\Bigg),
\end{equation}
\end{linenomath*}
where $\vec{\theta}$ is the set of nuisance parameters, $\hat{\mu}$ and $\hat{\vec{\theta}}$ are unconditional best fit values, and $\hat{\vec{\theta_{\mu}}}$ is the set of conditional best fit values of the nuisance parameters for a given value of $\mu$.
An upper limit on $\mu$ is determined using the profile likelihood statistic with the \CLs criterion.
The asymptotic approximation for the sampling distribution of $q(\mu)$ is assumed in the derivation of these results~\cite{cite:l1,cite:l2,cite:l3,Cowan:2010js}.
The expected significance under the SM hypothesis and the expected upper limits under the background-only hypothesis are also reported.
These are obtained by fitting to the corresponding Asimov data sets~\cite{Cowan:2010js}.

In addition, a combined maximum likelihood fit with the CMS measurement~\cite{CMS:2021kom} of \hgg using the same data sample is performed to determine the ratio \zgOvergg.
The \hgg analysis obtained a signal strength for $\sigppH\brgg$ of $1.12\pm0.09$.
In this combined fit, the branching fraction \brgg is an additional free parameter.
The uncertainty in the measured ratio of the two branching fractions is dominated by statistical uncertainty.
Common sources of theoretical and experimental uncertainty in the two measurements, described in the next section, are treated as correlated in the fit.
The combination is performed at $\hmass = \mH\GeV$, and the discrete profiling method is used for the background modeling in both cases. Although the event samples are predominantly independent, a small fraction of events containing the $\PZ(\ell\ell)\PH(\PGg\PGg)$ signature is selected in both analyses. It has been verified that this results in a negligible statistical correlation between the two measurements.

\section{Systematic uncertainties}
\label{sec:systematics}

The uncertainties associated with the choice of background shape are incorporated into the fit to the data through the use of the discrete profiling method.
They are, therefore, reflected in the statistical uncertainties obtained from the fit.
The systematic uncertainties, affecting either the normalization or the shape of the signal expectation, are listed below, and the numerical values are summarized in Table~\ref{tab:syst}, which also indicates whether the effect is correlated between the data-taking periods.
\begin{itemize}
  \item Theoretical cross section calculations: These include the effects of the choice of PDFs, the value of the strong coupling constant (\alpS), and the effect of missing higher orders in the perturbative cross section calculations, evaluated from variations of the renormalization and factorization scales (\muR, \muF)~\cite{cite:cs1,cite:cs2,Butterworth:2015oua}. The uncertainties are treated as independent for each Higgs boson production mechanism. The uncertainty in \brzg is also considered~\cite{LHC-YR4}.
  \item Underlying event and parton shower modeling: The uncertainty associated with the choice and tuning of the generator is estimated with dedicated samples which are generated by varying the parameters of the tune used to generate the original signal samples. The uncertainties are treated as correlated for the 2017 and 2018 samples, which use the CP5 tune~\cite{Sirunyan:2019dfx}, while being uncorrelated with the 2016 sample, which uses the CUETP8M1 tune~\cite{Khachatryan:2015pea}.
  \item Integrated luminosity: The integrated luminosities for the 2016, 2017, and 2018 data-taking years have uncertainties of 1.2\%, 2.3\%, and 2.5\%~\cite{CMS-LUM-17-003,LUM-17-004,LUM-18-002}, respectively, corresponding to an overall uncertainty for the 2016--2018 period of 1.6\%, the improvement in precision reflecting the (uncorrelated) time evolution of some systematic effects.
  \item L1 trigger: During the 2016 and 2017 data-taking periods, a gradual shift in the timing of the inputs of the ECAL L1 trigger in the $\abs{\eta} > 2.4$ region led to a specific inefficiency. A correction of approximately 1\% is applied to the simulation along with the corresponding uncertainty in the inefficiency measurement.
  \item Trigger: Uncertainties are evaluated for the corrections applied to the simulation to match the trigger efficiencies measured in data with \Ztoee and \Ztomm events.
  \item Photon identification and isolation: Uncertainties are evaluated for the corrections applied to the simulation to match the selection efficiencies in data measured with \Ztoee events.
  \item Lepton identification and isolation: Uncertainties are evaluated for the corrections applied to the simulation to match electron and muon selection efficiencies in data measured with \Ztoee and \Ztomm events.
  \item Pileup modeling: The uncertainty in the description of the pileup in the signal simulation is estimated by varying the total inelastic cross section by $\pm4.6$\%~\cite{Sirunyan:2018nqx}.
  \item Kinematic BDT: The uncertainties in the photon and lepton energy and the correction of the photon MVA discriminant are propagated to \Dkin. Changes in \Dkin cause the migration of signal events across category boundaries.
  \item VBF BDT: The uncertainties in the jet energy and the uncertainty in \Dkin are propagated to \DVBF. Changes in \DVBF cause the migration of signal events across category boundaries.
  \item Photon energy scale and resolution:
  The photon energy in the simulation is varied due to the ECAL energy scale and resolution uncertainties, and the effects on the signal mean and resolution parameters are propagated to the fits.
  \item Lepton momentum scale and resolution:
  The lepton momentum in the simulation is varied due to the lepton momentum scale and resolution uncertainties, and the effects on signal mean and resolution parameters are propagated to the fits.
\end{itemize}
In the \zgOvergg measurement, the common sources of theoretical and systematic uncertainty in the two analyses are treated as correlated in the fit.
These are the theoretical uncertainties in the Higgs production cross section calculations, and the systematic uncertainties in the underlying event and parton shower modeling, the integrated luminosity, and the L1 trigger inefficiency.
The remaining uncertainties are treated as uncorrelated.

\begin{table}[tbp]
  \centering
  \topcaption{Sources of systematic uncertainty affecting the simulated signal. The normalization effect on the expected yield, or the effect on the signal shape parameters, is given as indicated, with the values averaged over all event categories. The third column shows the uncertainties that have a correlated effect across the three data-taking periods.}
  \begin{tabular}{l@{\hskip 0.3in}c@{\hskip 0.3in}c}
  \hline
  Sources                &  Uncertainty (\%)   & Year-to-year correlation  \\ \hline
  \multicolumn{3}{c}{\textit{Normalization}}\\
  Theoretical &    &  \\
  -- \brzg & 5.7    &  Yes\\
  -- \ggH cross section (\muF, \muR) & 3.9 & Yes\\
  -- \ggH cross section (\alpS)& 2.6& Yes\\
  -- \ggH cross section (PDF)& 1.9 &Yes\\
  -- VBF cross section (\muF, \muR)&  0.4 & Yes\\
  -- VBF cross section (\alpS)&  0.5 & Yes\\
  -- VBF cross section (PDF)&  2.1 &Yes\\
  -- \WH cross section (\muF, \muR) & $^{+0.6}_{-0.7}$ &  Yes\\
  -- \WH cross section (PDF)& 1.7 & Yes\\
  -- \ZH cross section (\muF, \muR) & $^{+3.8}_{-3.1}$ &Yes \\
  -- \ZH cross section (PDF)& 1.3  &   Yes\\
  -- \WH/\ZH  cross section (\alpS) & 0.9 &Yes\\
  -- \ttH cross section (\muF, \muR)& $^{+5.8}_{-9.2}$ &  Yes\\
  -- \ttH cross section (\alpS)& 2.0 & Yes \\
  -- \ttH cross section (PDF)& 3.0 &Yes \\
  Underlying event and parton shower   	      & 3.7--4.4             &       	     Partial         \\
  Integrated luminosity  &  1.2--2.5 & Partial \\
  L1 trigger    	     &      	 	0.1--0.4      	 	   		 & No       \\
  Trigger                             						      &             	  &				  \\

  -- Electron  channel                  				                      				    &0.9--1.9	  &     No      \\
  -- Muon channel                       					                     					&0.1--0.4  	 &    No    \\
  Photon identification and isolation                      &    0.2--5.0            			 &  Yes           \\
  Lepton identification and isolation      &       	 	   & \\

  -- Electron channel                   			   	                     					&0.5--0.7  	 	 &      Yes      \\
  -- Muon channel                       				   	       					  			&0.3--0.4  	 &          Yes     \\
  Pileup                 				   			   & 0.4--1.0 & Yes            \\
  Kinematic BDT               & 2.5--3.7         				  &  Yes\\
  VBF BDT                            				 			   & 5.9--14.0 &  Yes\\
  \multicolumn{3}{c}{\textit{Shape parameters}}\\
  Photon energy and momentum					       &&             \\
  -- Signal mean                        						               					 & 0.1--0.4       &      Yes        \\
  -- Signal resolution                        						               		& 3.1--5.9       &      Yes        \\
  Lepton energy and momentum     				 			 				  	  &&          	                 \\
  -- Signal mean                        						       							 & 0.007  &     Yes   \\
  -- Signal resolution                  				  	     							   & 0.007--0.010 &     Yes          \\\hline
  \end{tabular}\label{tab:syst}
\end{table}

\section{Results}\label{sec:results}

\begin{figure*}[htbp]
\centering
\includegraphics[width=0.5\textwidth]{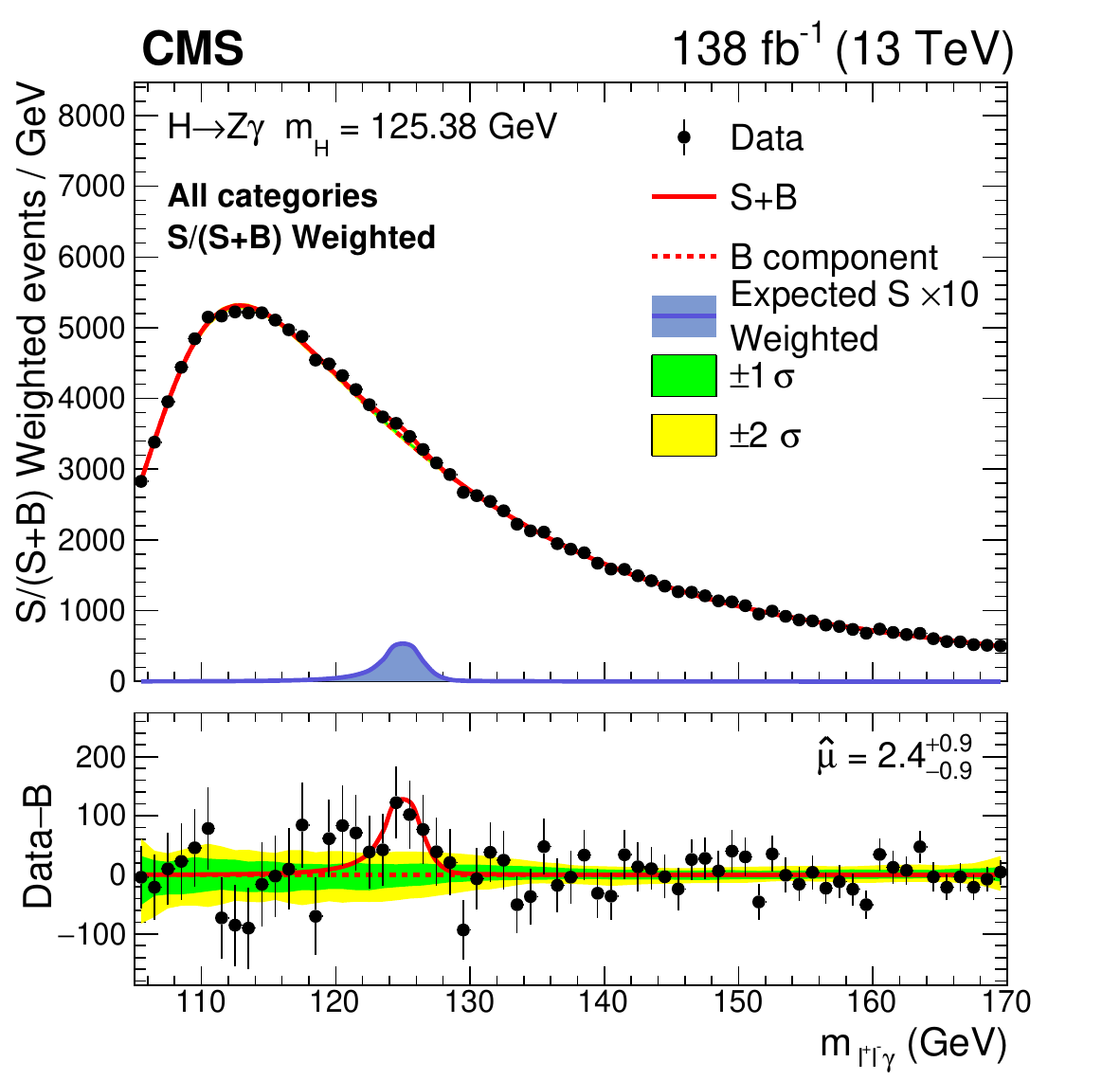}
 \caption{Sum over all categories of the data points and signal-plus-background model after the simultaneous fit to each \mllg distribution.
 The contribution from each category is weighted by $S/(S+B)$, as defined in the text.
 In the upper panel, the red solid line shows the signal-plus-background fit. The red dashed line shows the background component of the fit. The green and yellow bands represent the 68 and 95\% \CL uncertainties in the fit. Also plotted is the expected SM signal weighted by $S/(S+B)$ and scaled by a factor of 10. In the lower panel, the data minus the background component of the fit is shown.}\label{fig:SignalBackground}
\end{figure*}

Figure~\ref{fig:SignalBackground} shows the signal-plus-background fit to the data and the corresponding distribution after background subtraction for the sum of all categories.
Each category is weighted by the factor $S/(S+B)$, where $S$ is the measured signal yield and $B$ is the background yield in the narrowest mass interval containing 95\% of the signal distribution.

\begin{figure}[htbp]
  \centering
  \includegraphics[width=0.7\textwidth]{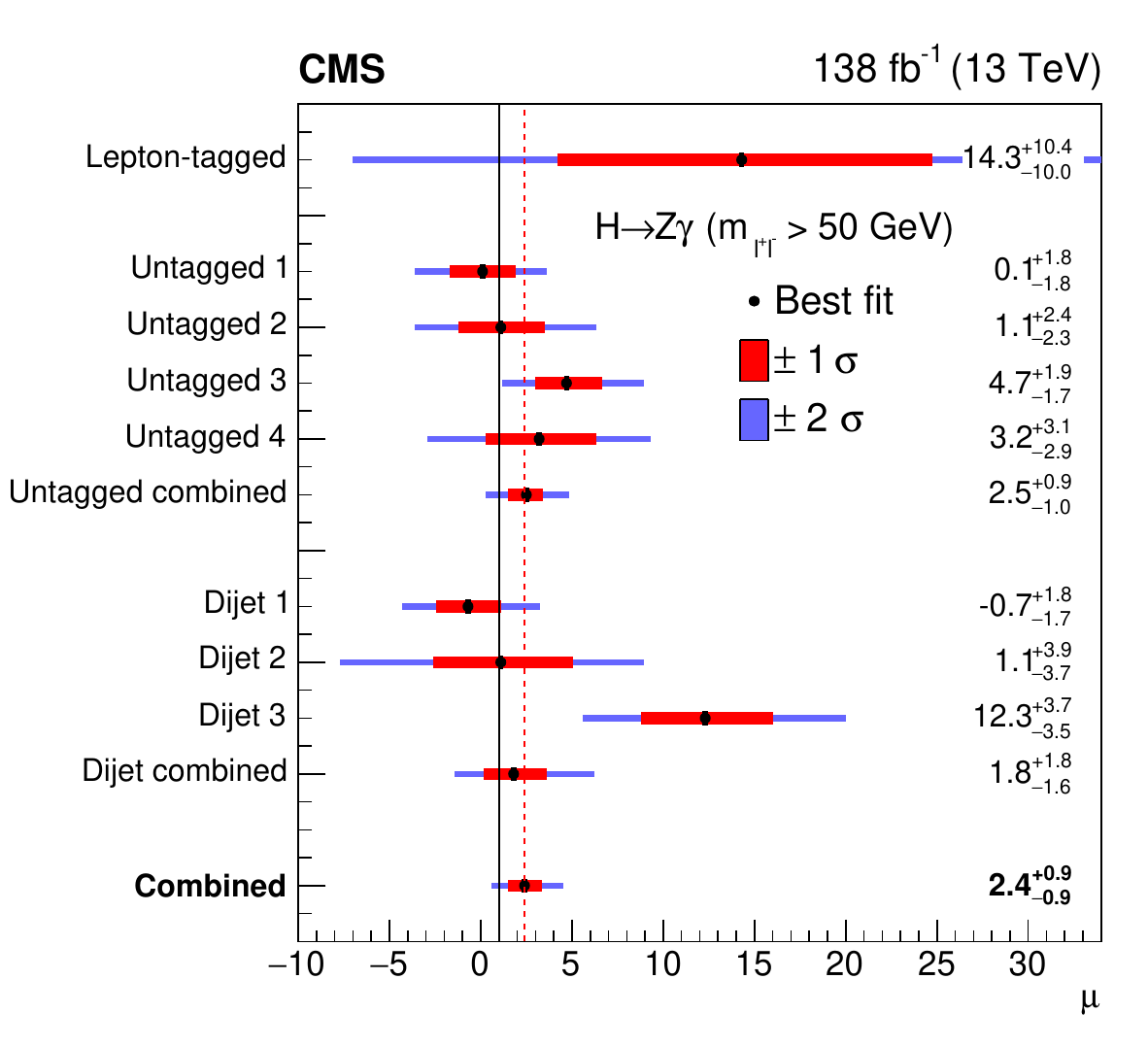}
  \caption{
Observed signal strength ($\mu$) for an SM Higgs boson with $\hmass=\mH\GeV$.
The labels ``untagged combined," ``dijet combined," and ``combined" represent the results obtained from simultaneous fits of the untagged categories, dijet categories, and full set of categories, respectively.
The black solid line shows $\mu=1$, and the red dashed line shows the best fit value $\hat{\mu}=\signalstrength$ of all categories combined.
The category compatibility $p$-value, described in the text, is \channelcompatp, corresponding to \channelcompatsigma standard deviations.}\label{fig:lim-combo125}
\end{figure}

\begin{figure}[htbp]
  \centering
  \includegraphics[width=0.7\textwidth]{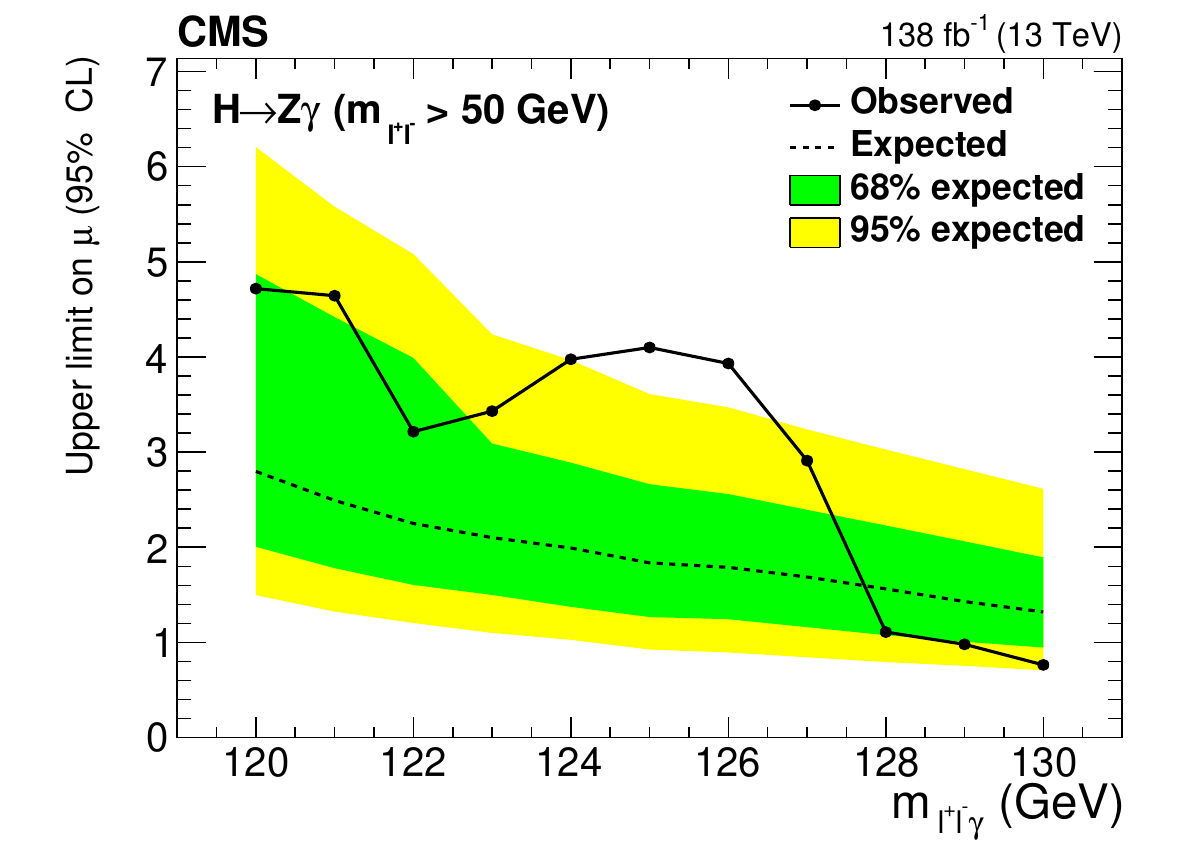}
    \caption{
      Upper limit (95\%~\CL) on the signal strength ($\mu$) relative to the SM prediction, as a function of the assumed value of the Higgs boson mass used in the fit.
    \label{fig:lim}}
\end{figure}

The best fit value of the signal strength is $\signalstrengthExpanded$ at $\hmass=\mH\GeV$. The corresponding measured value of \sigxbr is $\brExpanded\unit{pb}$.
This measurement is consistent with the SM prediction of $0.09 \pm 0.01\unit{pb}$ at the \compatibility standard deviation level.
Figure~\ref{fig:lim-combo125} shows the signal strengths obtained for each category separately, corresponding to the fit results shown in Figs.~\ref{fig:3} and~\ref{fig:4}, as well as from simultaneous fits to the dijet categories, the untagged categories, and all categories combined.
Among the eight categories, dijet 1 is the most sensitive.
A category compatibility $p$-value, under the hypothesis of a common signal strength in all categories, is calculated from the likelihood ratio between the nominal combined fit, in which all categories have the same signal strength parameter, and a separate fit, in which each category has its own signal strength parameter.
This $p$-value is found to be \channelcompatp, corresponding to \channelcompatsigma standard deviations, and is driven by the dijet 3 category, which has a signal strength of $\hat{\mu}=\signalstrengthdijetthree$.
The observed (expected) local significance is \obssig\,(\expsig) standard deviations.
Upper limits on $\mu$ are calculated at 1\GeV intervals in the mass range of $120 < \mllg < 130\GeV$ and at $\hmass=\mH\GeV$, as shown in Fig.~\ref{fig:lim}.
The observed (expected) limit at 95\% \CL relative to the SM prediction for $\hmass=\mH\GeV$ is \obslimit\,(\explimit), where the expected limit is calculated under the background-only hypothesis.
The measured value of \zgOvergg from the combined fit with the \hgg analysis is \brRatio.
This measurement is consistent with the SM prediction for the ratio at the \brRatioCompat standard deviation level.

\section{Summary}
\label{sec:summary}

A search is performed for a standard model (SM) Higgs boson decaying into a lepton pair (\epem or \mpmm) and a photon with $\mll>50\GeV$.
The analysis is performed using a sample of proton-proton (\pp) collision data at $\sqrt{s}=13\TeV$, corresponding to an integrated
luminosity of \LumiT\fbinv.
The main contribution to this final state is from Higgs boson decays to a \PZ boson and a photon ($\hzg\to\llg$).
The best fit value of the signal strength $\hat{\mu}$ for $\hmass=\mH\GeV$ is $\hat{\mu}=\signalstrengthExpanded=\signalstrength$.
This measurement corresponds to $\sigxbr=\br\unit{pb}$.
The measured value is \compatibility standard deviations higher than the SM prediction.
The observed (expected) local significance is \obssig\,(\expsig) standard deviations, where the expected significance is determined for the SM hypothesis.
The observed (expected) upper limit at 95\% confidence level on $\mu$ is \obslimit\,(\explimit).
In addition, a combined fit with the \hgg analysis of the same data set~\cite{CMS:2021kom} is performed to measure
the ratio $\zgOvergg=\brRatio$, which is consistent with the ratio of $0.69 \pm 0.04$ predicted by the SM at the \brRatioCompat standard deviation level.

\begin{acknowledgments}
  We congratulate our colleagues in the CERN accelerator departments for the excellent performance of the LHC and thank the technical and administrative staffs at CERN and at other CMS institutes for their contributions to the success of the CMS effort. In addition, we gratefully acknowledge the computing centers and personnel of the Worldwide LHC Computing Grid and other centers for delivering so effectively the computing infrastructure essential to our analyses. Finally, we acknowledge the enduring support for the construction and operation of the LHC, the CMS detector, and the supporting computing infrastructure provided by the following funding agencies: BMBWF and FWF (Austria); FNRS and FWO (Belgium); CNPq, CAPES, FAPERJ, FAPERGS, and FAPESP (Brazil); MES and BNSF (Bulgaria); CERN; CAS, MoST, and NSFC (China); MINCIENCIAS (Colombia); MSES and CSF (Croatia); RIF (Cyprus); SENESCYT (Ecuador); MoER, ERC PUT and ERDF (Estonia); Academy of Finland, MEC, and HIP (Finland); CEA and CNRS/IN2P3 (France); BMBF, DFG, and HGF (Germany); GSRI (Greece); NKFIH (Hungary); DAE and DST (India); IPM (Iran); SFI (Ireland); INFN (Italy); MSIP and NRF (Republic of Korea); MES (Latvia); LAS (Lithuania); MOE and UM (Malaysia); BUAP, CINVESTAV, CONACYT, LNS, SEP, and UASLP-FAI (Mexico); MOS (Montenegro); MBIE (New Zealand); PAEC (Pakistan); MES and NSC (Poland); FCT (Portugal); MESTD (Serbia); MCIN/AEI and PCTI (Spain); MOSTR (Sri Lanka); Swiss Funding Agencies (Switzerland); MST (Taipei); MHESI and NSTDA (Thailand); TUBITAK and TENMAK (Turkey); NASU (Ukraine); STFC (United Kingdom); DOE and NSF (USA).

\hyphenation{Rachada-pisek} Individuals have received support from the Marie-Curie program and the European Research Council and Horizon 2020 Grant, contract Nos.\ 675440, 724704, 752730, 758316, 765710, 824093, 884104, and COST Action CA16108 (European Union); the Leventis Foundation; the Alfred P.\ Sloan Foundation; the Alexander von Humboldt Foundation; the Belgian Federal Science Policy Office; the Fonds pour la Formation \`a la Recherche dans l'Industrie et dans l'Agriculture (FRIA-Belgium); the Agentschap voor Innovatie door Wetenschap en Technologie (IWT-Belgium); the F.R.S.-FNRS and FWO (Belgium) under the ``Excellence of Science -- EOS" -- be.h project n.\ 30820817; the Beijing Municipal Science \& Technology Commission, No. Z191100007219010; the Ministry of Education, Youth and Sports (MEYS) of the Czech Republic; the Hellenic Foundation for Research and Innovation (HFRI), Project Number 2288 (Greece); the Deutsche Forschungsgemeinschaft (DFG), under Germany's Excellence Strategy -- EXC 2121 ``Quantum Universe" -- 390833306, and under project number 400140256 - GRK2497; the Hungarian Academy of Sciences, the New National Excellence Program - \'UNKP, the NKFIH research grants K 124845, K 124850, K 128713, K 128786, K 129058, K 131991, K 133046, K 138136, K 143460, K 143477, 2020-2.2.1-ED-2021-00181, and TKP2021-NKTA-64 (Hungary); the Council of Science and Industrial Research, India; the Latvian Council of Science; the Ministry of Education and Science, project no. 2022/WK/14, and the National Science Center, contracts Opus 2021/41/B/ST2/01369 and 2021/43/B/ST2/01552 (Poland); the Funda\c{c}\~ao para a Ci\^encia e a Tecnologia, grant CEECIND/01334/2018 (Portugal); MCIN/AEI/10.13039/501100011033, ERDF ``a way of making Europe", and the Programa Estatal de Fomento de la Investigaci{\'o}n Cient{\'i}fica y T{\'e}cnica de Excelencia Mar\'{\i}a de Maeztu, grant MDM-2017-0765 and Programa Severo Ochoa del Principado de Asturias (Spain); the Chulalongkorn Academic into Its 2nd Century Project Advancement Project, and the National Science, Research and Innovation Fund via the Program Management Unit for Human Resources \& Institutional Development, Research and Innovation, grant B05F650021 (Thailand); the Kavli Foundation; the Nvidia Corporation; the SuperMicro Corporation; the Welch Foundation, contract C-1845; and the Weston Havens Foundation (USA).
\end{acknowledgments}

\bibliography{auto_generated}
\cleardoublepage \appendix\section{The CMS Collaboration \label{app:collab}}\begin{sloppypar}\hyphenpenalty=5000\widowpenalty=500\clubpenalty=5000
\cmsinstitute{Yerevan Physics Institute, Yerevan, Armenia}
{\tolerance=6000
A.~Tumasyan\cmsorcid{0009-0000-0684-6742}
\par}
\cmsinstitute{Institut f\"{u}r Hochenergiephysik, Vienna, Austria}
{\tolerance=6000
W.~Adam\cmsorcid{0000-0001-9099-4341}, J.W.~Andrejkovic, T.~Bergauer\cmsorcid{0000-0002-5786-0293}, S.~Chatterjee\cmsorcid{0000-0003-2660-0349}, K.~Damanakis\cmsorcid{0000-0001-5389-2872}, M.~Dragicevic\cmsorcid{0000-0003-1967-6783}, A.~Escalante~Del~Valle\cmsorcid{0000-0002-9702-6359}, R.~Fr\"{u}hwirth\cmsAuthorMark{1}\cmsorcid{0000-0002-0054-3369}, M.~Jeitler\cmsAuthorMark{1}\cmsorcid{0000-0002-5141-9560}, N.~Krammer\cmsorcid{0000-0002-0548-0985}, L.~Lechner\cmsorcid{0000-0002-3065-1141}, D.~Liko\cmsorcid{0000-0002-3380-473X}, I.~Mikulec\cmsorcid{0000-0003-0385-2746}, P.~Paulitsch, F.M.~Pitters, J.~Schieck\cmsAuthorMark{1}\cmsorcid{0000-0002-1058-8093}, R.~Sch\"{o}fbeck\cmsorcid{0000-0002-2332-8784}, D.~Schwarz\cmsorcid{0000-0002-3821-7331}, S.~Templ\cmsorcid{0000-0003-3137-5692}, W.~Waltenberger\cmsorcid{0000-0002-6215-7228}, C.-E.~Wulz\cmsAuthorMark{1}\cmsorcid{0000-0001-9226-5812}
% ...
\par}
\vskip\cmsinstskip
\dag:~Deceased\\
$^{1}$Also at TU Wien, Vienna, Austria\\
% ...
\end{sloppypar}
%%% END EDITABLE REGION %%%
% skeleton_end
\end{document}